\documentclass[fleqn,10pt]{wlscirep}

\usepackage{verbatim}
\usepackage{dsfont}
\usepackage{cleveref}
\usepackage{mathpazo}

\newcommand{\bra}[1]{\left\langle #1\right|}
\newcommand{\ket}[1]{\left| #1\right\rangle}

\newcommand{\beq}{\begin{equation}}
\newcommand{\eeq}{\end{equation}}

\newcommand{\tr}{\text{Tr}}

\title{The quantum Zeno and anti-Zeno effects with driving fields in the weak and strong coupling regimes}

\author[1]{Mehwish Majeed}
\author[2,*]{Adam Zaman Chaudhry}
\affil[1,2]{School of Science \& Engineering, Lahore University of Management Sciences (LUMS), Opposite Sector U, D.H.A, Lahore 54792, Pakistan}

\affil[*]{adam.zaman@lums.edu.pk}

\begin{abstract}

Repeated measurements in quantum mechanics can freeze (the quantum Zeno effect) or enhance (the quantum anti-Zeno effect) the time-evolution of a quantum system. In this paper, we present a general treatment of the quantum Zeno and anti-Zeno effects for arbitrary driven open quantum systems, assuming only that the system-environment coupling is weak. In particular, we obtain a general expression for the effective decay rate of a two-level system subjected to arbitrary driving fields as well as periodic measurements. We demonstrate that the driving fields change the decay rate, and hence the quantum Zeno and anti-Zeno behavior, both qualitatively and quantitatively. We also extend our results to systems consisting of more than one two-level system, as well as a two-level system strongly coupled to an environment of harmonic oscillators, to further illustrate the non-trivial effect of the driving fields on the quantum Zeno and anti-Zeno effects. 
\end{abstract}

\begin{document}

\flushbottom
\maketitle
%
%
\thispagestyle{empty}

\section{Introduction}

Unlike classical measurements, quantum measurements in general disturb the state of the quantum system.  This back-action of measurements is a peculiar concept of quantum
mechanics which gives rise to striking phenomena such as the quantum Zeno effect (QZE).
In the QZE, repeated measurements hinder the time evolution of the quantum system\cite{Sudarshan1977,FacchiPhysLettA2000,FacchiPRL2002,
FacchiJPA2008,WangPRA2008,ManiscalcoPRL2008,FacchiJPA2010,MilitelloPRA2011,RaimondPRA2012,SmerziPRL2012,
WangPRL2013,McCuskerPRL2013,StannigelPRL2014,ZhuPRL2014,SchafferNatCommun2014,SignolesNaturePhysics2014,
DebierrePRA2015,AlexanderPRA2015,QiuSciRep2015,HePRA2018,HanggiNJP2018,HePRA2019,MullerAnnPhys}. However, it has also been observed that if the measurements are not rapid enough, a reverse effect, known as a quantum anti-Zeno effect (QAZE), can occur whereby the measurements accelerate the
quantum evolution \cite{KurizkiNature2000,RaizenPRL2001,BaronePRL2004,KoshinoPhysRep2005,BennettPRB2010,YamamotoPRA2010,ChaudhryPRA2014zeno,Chaudhryscirep2017b,HePRA2017,WuPRA2017,Chaudhryscirep2018,WuAnnals2018,ChaudhryEJPD2019a}. Both the QZE
and the QAZE have gained considerable interest theoretically and experimentally due
to their huge importance in the foundations of quantum mechanics as well as possible
applications in quantum technologies. For example, the effect of repeated measurements can be used to infer properties of the environment of a quantum system, that is, noise sensing \cite{sakuldee2020,mullerPLA2020,sakuldeePRA2020}. Generally speaking, so far, the main focus of the studies performed on QZE and the QAZE have been on the population decay model \cite{KurizkiNature2000,RaizenPRL2001,BaronePRL2004,KoshinoPhysRep2005,ManiscalcoPRL2006,SegalPRA2007,ZhengPRL2008,BennettPRB2010,YamamotoPRA2010,AiPRA2010,ThilagamJMP2010,ThilagamJCP2013} and the dephasing model \cite{ChaudhryPRA2014zeno} (notable exceptions include Refs.~{\renewcommand{\citemid}{}\cite[]{Chaudhryscirep2016,Chaudhryscirep2017a}). 

What is lacking is a rigorous general study of the QZE and QAZE in the presence of coherent driving fields. The idea of controlling the coherent dynamics of a quantum system by an external time-dependent field has found
widespread theoretical and experimental interest in many areas of physics and chemistry \cite{Hanggidrivenquantumtunneling}. For example, driving fields are a commonly used tool to manipulate qubits as well as to control chemical reactions by external laser fields \cite{kofmanPRL2001,kofmanPRL2004,gordonJPB2007,gordonPRL2008}. In quantum optics, it has been shown that
a frequency-modulated excitation of a two-level atom significantly modifies the time-evolution of the system \cite{noelPRA1998} .  In quantum tunneling systems, it has been demonstrated that an appropriately designed coherent drive can bring the tunneling to an (almost) complete standstill - this is known as coherent destruction of tunneling \cite{grossmannPRL1991,shaoPRA1997}. Driving fields can even effectively remove the interaction of the quantum system with its environment, which is precisely the idea behind dynamical decoupling \cite{ViolaPRA1998,LloydPRL1999,FanchiniPRA12007,ChaudhryPRA12012,ChaudhryPRA22012,ChaudhryPRA2019}. Driving fields have also been recently used in noise sensing \cite{doNJP2019}.

It is then clear that driving fields, repeated measurements as well as the environment can all drastically influence the temporal evolution of a quantum system. Consequently, in this work, we study the QZE and the QAZE when driving fields are applied to the quantum system as well. We start by deriving a general expression of the effective decay rate for the driven quantum system, provided that the system-environment coupling is weak, thereby extending the formalism of Ref.~{\renewcommand{\citemid}{}\cite[]{Chaudhryscirep2016}. We then obtain general expressions for the decay rate of a two-level system subjected to arbitrary driving fields. In particular, we consider in detail both the population decay model and the pure dephasing model in the presence of different driving fields. For example, we show that the effective decay rate for the driven population decay model can no longer be obtained using the usual sinc-squared function (as can be done in the absence of any driving fields \cite{KurizkiNature2000}). Moreover, counter-rotating terms in the system-environment interaction Hamiltonian can become important in the presence of the driving fields, in contrast with the undriven case. We then extend our results to more than one two-level system by modeling the multiple two-level systems as a single large spin \cite{VorrathPRL2005}. We also demonstrate that our results can be extended to the strong system-environment coupling regime via the well-known polaron transformation  \cite{SilbeyJCP1984,Vorraththesis,LeeJCP2012,changJCP2013,jang2008theory,ChinPRL2011,GuzikJPCL2015} along with perturbation theory. All in all, our results generally indicate that the effective decay rate is very significantly influenced by the driving fields.

\section{Results}

\subsection*{Effective rate of an arbitrary driven quantum system in the weak coupling regime}
We start by writing the total Hamiltonian of a quantum system, in the presence of driving fields, interacting with its environment as
\begin{equation}\label{D1}
\hat{H}(t) = \hat{H}_S(t) + \hat{H}_B + \hat{H}_{SB}. 
\end{equation}
Here, the first term $\hat{H}_S(t)$ describes  the central quantum system Hamiltonian. This carries explicit time-dependence due to the application of external driving fields on the system; consequently, we write it as the sum of a time-independent part $\hat{H}_S$ and a time-dependent part $\hat{H}_c(t)$ describing the effect of the external fields. The second term $\hat{H}_B$ corresponds to environment, whereas the last term $\hat{H}_{SB}$ is  the coupling between them, which, for later convenience, we write in the diagonal form $\hat{H}_{SB} = \sum_\mu \hat{F}_\mu \otimes \hat{B}_\mu$, with the $\hat{F}_\mu$ operators belonging to the system Hilbert space and the $\hat{B}_\mu$ operators living in the environment Hilbert space. From now on, we will be suppressing the `hats' on the operators - the context should make it clear whether or not we are dealing with an operator. Keeping in mind our objective of investigating the quantum Zeno and anti-Zeno effects in such driven open quantum systems, our primary quantity of interest is the effective decay rate of the system when repeated projective measurements are performed on the system with time interval $\tau$. To calculate the effective decay rate, we assume that the system is initially prepared in the pure state $\ket{\psi}$. We then find the system density matrix  at time $\tau$, that is, $\rho_S(\tau)$, and then use this to find the survival probability $s(\tau)$ that the system is still in state $\ket{\psi}$ at time $\tau$. Thereafter, we can find the effective decay rate $\Gamma(\tau)$ via $\Gamma(\tau)=-\text{ln}\,s(\tau)/\tau$. The system density matrix at time $\tau$ is obtained via $\rho_S(\tau)=\tr_B[U(\tau)\rho(0)U^{\dagger}(\tau)]$, where $\rho(0)$ is the state of total system plus environment, $\tr_B$ is the partial trace with respect to  environment states and $U(\tau)$ is the total unitary time-evolution operator corresponding to the total Hamiltonian $H(t)$. Generally speaking, for the time-dependent system-environment models considered here, it is usually impossible to calculate the  time-evolution operator exactly. However, for weakly coupled system-environment models, we can find the time-evolution operator $U(\tau)$ using time-dependent perturbation theory \cite{Sakuraibook}. We assume that the system-environment state is initially of a simple product form, that is, $\rho(0)=\ket{\psi}\bra{\psi}\otimes\rho_B$,  where $\rho_B=e^{-\beta H_B}/Z_B$ is the thermal equilibrium state of the environment with $Z_B=\tr_B[e^{-\beta H_B}]$. Extending the treatment of Ref.~{\renewcommand{\citemid}{}\cite[]{Chaudhryscirep2016} to time-dependent Hamiltonians in a straightforward manner, we find that the decay rate of the quantum state $\ket{\psi}$, in the presence of projective measurements and for weak system-environment coupling, is given by `overlap integral' of two functions - the generalized filter function $Q(\omega,\tau)$ and the spectral density of the environment $J(\omega)$ (see the Methods section), that is,  
\begin{equation}\label{Decayrate}
\Gamma(\tau) = \int_0^\infty \, d\omega \,Q(\omega,\tau) J(\omega).
\end{equation}\\
Here, the generalized filter function is given by 
\begin{equation}\label{Filterfunction}
Q(\omega,\tau) =\dfrac{2}{\tau}\text{Re}\left(
\sum_{\mu \nu}\int_0^\tau dt \int_0^{t} dt'  f_{\mu \nu}(\omega,t') 
\tr_S\bigr[P_{\perp}\widetilde{F}_\nu(t-t')\rho_S(0)\widetilde{F}_\mu(t) \bigr]\right),
\end{equation}\\
where $F_\mu(t) = U_S^\dagger (t) H_S(t) U_S(t)$, with $U_S(t)$ being the unitary time-evolution operators corresponding to $H_S(t)$ only, and $P_{\perp}$ is the projector onto the system subspace orthogonal to $\ket{\psi}\bra{\psi}$. The environment correlation function is $C_{\mu\nu}(t)  = \tr_B [\rho_B e^{iH_B t} B_\mu e^{-iH_B t} B_\nu]$, which can generally be simplified to the form $C_{\mu\nu}(t) = \sum_k |g_k|^2f_{\mu\nu}(\omega_k,t)$, where $g_k$ is the
coupling between the system and the $k^{\text{th}}$ mode of the environment. The function $f_{\mu\nu}(\omega_k,t)$ then contains the remaining information about $C_{\mu\nu}(t)$. The sum over the modes is typically converted to an integral over the environment frequencies via the substitution $\sum_k
|g_k|^2(\hdots) \rightarrow \int_0^\infty \, d\omega \, J(\omega) (\hdots.)$, thereby introducing the spectral density function $J(\omega)$ of the environment. It should be noted that $Q(\omega,\tau)$ depends not only on the frequency of the measurements, the way that system is coupled to its environment, the state of system that is repeatedly prepared, and part of the environment correlation function $f_{\mu \nu}(\omega,t)$; most importantly for us, it also depends on the driving fields applied. Similar analytical expressions to account for the effect of driving fields have been considered before \cite{kofmanPRL2001,kofmanPRL2004,gordonJPB2007,gordonPRL2008}. However, our expression takes into account the effect of both measurements as well as the concurrent application of driving fields for arbitrary system-environment models, and we do not make any assumptions regarding the driving fields such as the adiabatic approximation \cite{MilitelloPRA2019a,MilitelloPRA2019b}. We also note that there are different ways to define the survival probability and hence the decay rate, as well as different ways of identifying the Zeno and anti-Zeno regimes. For example, one can also look at the history of measurements \cite{Halliwellhistoriesreview, DankoPRA2018} when calculating the survival probability \cite{Chaudhryscirep2018}. Similarly, we identify the Zeno and anti-Zeno regimes by looking at when the decay rate $\Gamma(\tau)$ is an increasing function (the Zeno regime) or a decreasing function (the anti-Zeno regime) \cite{Chaudhryscirep2016}; an alternative approach is to compare the measurement modified decay rate with the decay rate without measurement \cite{FacchiPRL2001}.

\subsection*{General expression of the decay rate for a driven two-level system}
To apply our formalism to a two-level system, we first note that, without loss of generality, we can assume the initial state to be $\ket{e}$, where $\sigma_z\ket{e}=\ket{e}$, since we can always choose our coordinate system in this manner. We then check, with time interval $\tau$, whether or not the system is still in this state or not. The projector onto the orthogonal subspace is $\ket{g}\bra{g}$, where $\sigma_z \ket{g} = -\ket{g}$. Now comes a key insight. No matter what the external driving fields are, the system unitary time-evolution operator corresponding to $U_S(t)$ can always be written in the form 
\begin{equation}\label{Er1}
U_S(t)=e^{-i \alpha(t) \sigma_z/2}e^{-i \beta(t) \sigma_y/2}e^{-i\gamma(t) \sigma_z/2},
\end{equation}
where $\alpha(t)$, $\beta(t)$, and $\gamma(t)$ are time-dependent Euler angles. These are arbitrary functions of time with the constraint that $U_S(t = 0) = \mathds{1}$. The corresponding Hamiltonian $H_S(t)$ can be worked out from Schrodinger's equation. We find that 
\begin{align}\label{EH}
H_S(t)&=\dfrac{\sigma_z}{2}\dfrac{\partial \alpha(t)}{\partial t}+\dfrac{1}{2}\dfrac{\partial \beta(t)}{\partial t}\bigr(\cos[\alpha(t)]\sigma_y-\sin[\alpha(t)]\sigma_x\bigr)+\dfrac{1}{2}\dfrac{\partial \gamma(t)}{\partial t}\cos[\beta(t)]\sigma_z \, \notag\\
&+\dfrac{1}{2}\dfrac{\partial \gamma(t)}{\partial t}\bigr(\sin[\beta(t)]\cos[\alpha(t)]\sigma_x+\sin[\beta(t)]\sin[\alpha(t)]\sigma_y\bigr).
\end{align}
In other words, by choosing the functions $\alpha(t)$, $\beta(t)$, and $\gamma(t)$ appropriately, we can work backwards to find the corresponding system Hamiltonian. 

With this form of $U_S(t)$, we can work out the effective decay rate. Using our previous general expression given in Eq.~\eqref{Filterfunction}, we find that the filter function $Q(\omega,\tau)$ is 
\begin{equation}\label{filter}
Q(\omega,\tau) =\frac{2}{\tau}\text{Re} \left(\sum_{\mu\nu}\int_0^\tau dt \int_0^t dt'f_{\mu\nu}(\omega,t')G_\mu(t)\bar{G}_\nu(t-t')\right),
\end{equation}
where 
\begin{align}\label{Gmu}
G_\mu(t) = e^{i(\alpha(t)+\gamma(t))}\cos^2[\dfrac{\beta(t)}{2}]F_{\mu eg}-e^{-i(\alpha(t)-\gamma(t))}\sin^2[\dfrac{\beta(t)}{2}]F_{\mu ge}+e^{i\gamma(t)}\dfrac{\sin[\beta(t)]}{2}(F_{\mu gg}-F_{\mu ee}),
\end{align}
and 
\begin{align}\label{Gnu}
\bar{G}_\nu(t) =e^{-i(\alpha(t)+\gamma(t))}\cos^2[\dfrac{\beta(t)}{2}]F_{\nu ge}-e^{i(\alpha(t)-\gamma(t))}\sin^2[\dfrac{\beta(t)}{2}]F_{\nu eg}+e^{-i\gamma(t)}\dfrac{\sin[\beta(t)]}{2}(F_{\nu gg}-F_{\nu ee}),
\end{align}
with $F_{\mu ij}=\bra{i}F_{\mu}\ket{j}$. In the following sections, we use this form of the filter function with various $\alpha(t)$, $\beta(t)$, and $\gamma(t)$ to evaluate the effective decay rate for different system-environment models.

\subsection*{Application to the driven population decay model in weak coupling regime}

To illustrate our formalism, we first consider a single two-level system interacting with an environment of harmonic oscillators in the presence of external driving fields. The total system-environment Hamiltonian is written as (we set $\hbar=1$ throughout)
\begin{equation}\label{CT}
H(t) = \frac{\varepsilon_0}{2}\sigma_z +H_c(t)+ \sum_k \omega_k b_k^\dagger b_k + \sum_k (g_k^*b_k \sigma_+ + g_k b_k^\dagger \sigma_-),
\end{equation}
where $H_{S}(t)=\frac{\varepsilon_0}{2}\sigma_z+H_c(t)$ is the system Hamiltonian with $\varepsilon_0$  representing  the energy spacing of the two-level system, while $H_c(t)$ is a time-dependent external driving field acting on the system.  $H_{B} =\sum_k \omega_k b_k^\dagger b_k$ is the environment Hamiltonian with $b_k$ and $b_k^\dagger$ representing the usual annihilation and creation operators, and $H_{SB} =\sum_k (g_k^*b_k \sigma_+ + g_k b_k^\dagger \sigma_-)$ is the system-environment interaction Hamiltonian with $g_k$ denoting the coupling strength between the central two-level system and the environment oscillators. As usual, $\sigma_z$  is the standard Pauli spin-1/2 matrix, and $\sigma_+(\sigma_-)$ are the raising (lowering)  operators. Note that we have made the rotating-wave approximation (RWA) here for the system-environment interaction, which means that we have neglected those processes which do not conserve energy \cite{DebierrePRA2015,VegaRMP2017}.

\begin{figure}[t!]
	\includegraphics[scale = 0.85]{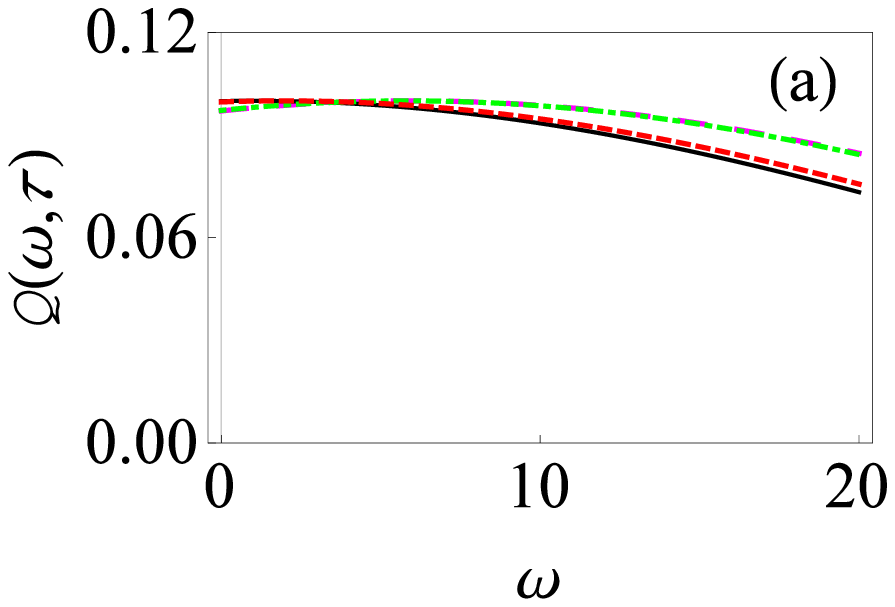}
	\quad
	\includegraphics[scale = 0.85]{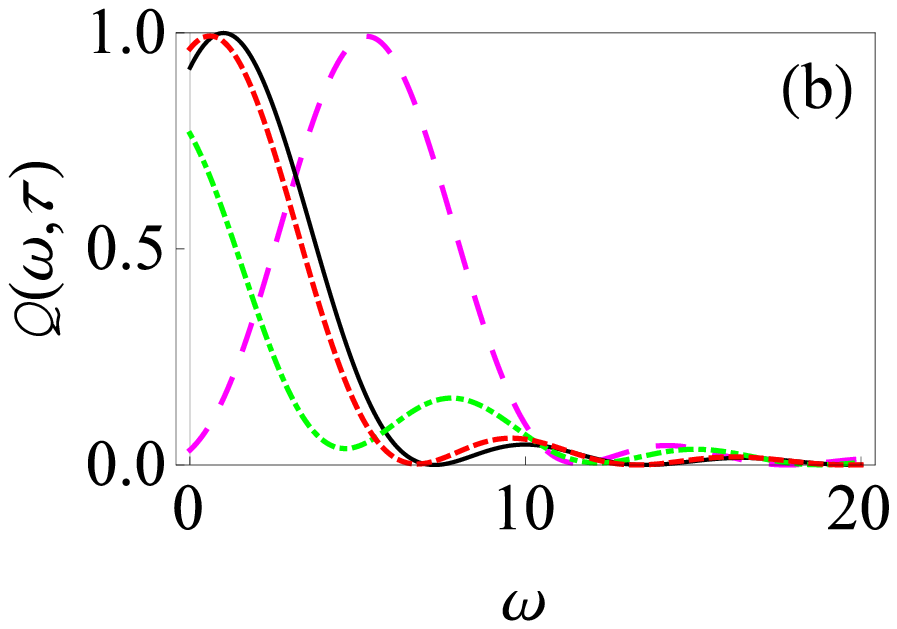}
	
	\centering
	\caption{\textbf{Filter function for the driven population decay model with the RWA}. $\textbf{(a)}$  Graph of  $Q(\omega,\tau)$ versus  $\omega$  in the presence of time-dependent driving fields: $\alpha(t)=\varepsilon_0 t +( V_0/\Omega) \sin(\Omega t)$,  $\beta(t)=0$ and $\gamma(t)=0$  with $V_0=1$ and $\Omega=5$ (dashed, red curve), $V_0=5$ and $\Omega=1$ (large-dashed, magenta curve), and $V_0=5$ and $\Omega=5$ (dot-dashed, green curve). $\alpha(\tau)=\varepsilon_0 \tau$, $\beta(\tau)=0$ and $\gamma(\tau)=0$ for the solid, black curve. Throughout, we use dimensionless units with $\hbar = 1$ and we have set $\varepsilon_0 = 1$. Here we have used a relatively small measurement interval, that is, $\tau=0.1$.  $\textbf{(b)}$  Same as $\textbf{(a)}$, except that now $\tau=1$.}
	\label{Qab}
\end{figure}

With the model specified, we now move to find the effective decay rate of the two-level system using the formalism described before. As is usually the case in studies of the quantum Zeno and anti-Zeno effects, we initially prepare our system in the excited state $\ket{e}$ such that $\sigma_z\ket{e}=\ket{e}$, and we then repeatedly check, with time interval $\tau$, whether or not the system is still in the excited state. To calculate the decay rate using our formalism, we note that $F_1 = \sigma_+$, $F_2 =\sigma_-$, $C_{\mu\nu}(t)=\tr_B[\rho_B\widetilde{B}_\mu(t)B_\nu]$, $\widetilde{B}_\mu(t)=e^{iH_Bt}B_\mu e^{-iH_Bt}$, $B_1 = \sum_k g_k^* b_k$, and $B_2 = \sum_k g_k b_k^\dagger$. In the limit of zero temperature, we find that $f_{12}(\omega,t) = e^{-i\omega t}$, while $f_{11} = f_{22} = f_{21} = 0$. Moreover, we find $
G_1(t)=e^{i(\alpha(t)+\gamma(t))} \cos^2[\beta(t)/2] $,
and
$\bar{G}_2(t-t')= e^{-i(\alpha(t-t')+\gamma(t-t'))} \cos^2[\beta(t-t')/2]$. Using these results, we obtain 
\begin{equation}\label{rotating}
Q(\omega,\tau) =\frac{2}{\tau} \int_0^\tau dt \int_0^t dt'\cos[\alpha(t)-\alpha(t-t')+\gamma(t)-\gamma(t-t')-\omega t'] \cos^2[\beta(t)/2] \cos^2[\beta(t-t')/2].
\end{equation}\\
In general, this can be a very complicated function. Therefore, we first consider the simplest case where $\alpha(t)=\varepsilon_0 t$, while $\beta(t) = \gamma(t) = 0$. This corresponds to [see Eq.~\eqref{EH}] $H = \frac{\varepsilon_0}{2}\sigma_z$, that is, the usual population decay model with no driving field. After performing the integrals, we get  $Q(\omega,\tau)=\tau \text{sinc}^2[\dfrac{(\varepsilon_0-\omega)\tau}{2}]$, thereby reproducing the well-known sinc-squared  function for $Q(\omega,\tau)$. Our formalism, on the other hand, allows us to go much further. The next case that we can consider is $\alpha(t)=\varepsilon_0 t+(V_0/\Omega)\sin(\Omega t)$ with $\beta(t) = \gamma(t) = 0$, which corresponds to $H_S = \frac{\varepsilon_0}{2}\sigma_z$ and $H_c(t) = V_0 \cos(\Omega t)\sigma_z/2$, with $V_0$ the amplitude of the applied sinusoidal field and $\Omega$ its frequency. Using the Jacobi-Auger identity $e^{ix \sin(\Omega t)}=\sum_{l=-\infty}^{\infty}J_l(x)e^{il\Omega t}$, with $J_l(x)$ being the Bessel functions of the first kind \cite{Gradshteynbook}, we find that now  
\begin{equation}\label{Bessel}
Q(\omega,\tau)=\sum_{m,n=-\infty}^{\infty}\frac{\tau}{\varepsilon_0-\omega+m\Omega}A_{mn}\bigr(\text{sinc}^2[\dfrac{(m-n)\Omega\tau}{2}](m-n)\Omega+\text{sinc}^2[\dfrac{(\varepsilon_0-\omega+n\Omega)\tau}{2}](\varepsilon_0-\omega+n\Omega)\bigr),
\end{equation}
with $A_{mn}=J_m(V_0/\Omega)J_n(V_0/\Omega)$. It is clear that the filter function is no longer a simple sinc-squared function - although the average value of the sinusoidal applied field is zero, the filter function changes in a very non-trivial manner. In particular, it is clear that the filter function $Q(\omega, \tau)$ is no longer generally peaked at $\omega = \epsilon_0$, even for changing measurement interval $\tau$. Rather, the second term in Eq.~\eqref{Bessel} makes it particularly clear that such a simple conclusion no longer holds in the driven case, and in fact the peak of the filter function changes as the measurement interval changes. Carrying on, we can also consider $\alpha(t)=\varepsilon_0 t$, with non-zero $\beta(t)$ (while $\gamma(t) = 0$). This corresponds to the driving field $H_c(t)=\dfrac{1}{2}\dfrac{\partial \beta(t)}{\partial t}\bigr(\cos[\varepsilon_0 t]\sigma_y-\sin[\varepsilon_0 t]\sigma_x\bigr)$. In these cases, $Q(\omega,\tau)$ needs to be calculated numerically, but the point is that in all such cases, the filter function, and hence the decay rate, changes in a very non-trivial manner. Similarly, we can also study non-zero values of $\gamma(t)$; the part of the Hamiltonian which contributes in $H_c(t)$ due to $\gamma(t)$ is of form $\dfrac{1}{2}\dfrac{\partial \gamma(t)}{\partial t}\cos[\beta(t)]\sigma_z+\dfrac{1}{2}\dfrac{\partial \gamma(t)}{\partial t}\bigr(\sin[\beta(t)]\cos[\alpha(t)]\sigma_x+\sin[\beta(t)]\sin[\alpha(t)]\sigma_y\bigr)$.

We now illustrate the change in the filter function as a result of these driving fields. Our results are shown in Fig. \ref{Qab} where we demonstrate the behavior of the filter function $Q(\omega,\tau)$ as a function of oscillator frequency $\omega$  for two different measurement intervals $\tau$ both with and without driving fields. We have set $\beta(t) = \gamma(t) = 0$, while $\alpha(t) = \varepsilon_0 t$ for the solid, black curve (the undriven case) and $\alpha(t) = \varepsilon_0 t +( V_0/\Omega) \sin(\Omega t)$ for the other curves (the driven cases). For the small measurement interval case illustrated in Fig.~\ref{Qab}(a), the different filter functions practically overlap - this is simply a manifestation of the convergence to the Zeno limit in the small measurement interval scenario even in the presence of the driving fields. However, for relatively large measurement interval $\tau$, the filter function for the population decay model (given by the usual sinc-squared function) is qualitatively different from the cases where we place the central system in a time-dependent external field  [see Fig.~\ref{Qab}(b)]. It can be seen that for the solid, black curve (the no driving case), the filter function $Q(\omega, \tau)$ is sharply peaked at $\varepsilon_0=\omega$ for $\tau=1$, and  changes very appreciably in the presence of strong driving fields (the dot-dashed green and long-dashed magenta curves). The long-dashed magenta curve corresponds to relatively lower frequency ($V_0 = 5$ and $\Omega = 1$), and to a first approximation, this filter function can be obtained by considering that the peak of the usual sinc-squared filter function is now shifted to $\varepsilon_0 + V_0$. However, for strong driving fields with higher frequencies (the dot-dashed green curve), such a naive picture is no longer applicable. Looking at Eq.~\eqref{Bessel} and using the fact that for higher frequencies, the Bessel functions are rapidly decaying so that only a few terms in the sum are important, it is clear that not only is the frequency $\omega = \varepsilon_0$ important in the filter function, but also other frequencies such as $\omega = \varepsilon_0 + \Omega$, $\omega = \varepsilon_0 - \Omega$, and so on. This leads to a much richer and complicated filter function, whose peak in fact also changes as the measurement interval $\tau$ is varied. As a result, we can expect that the effective decay rate is non trivially modified.

\begin{figure}[t!]
	\includegraphics[scale = 0.85]{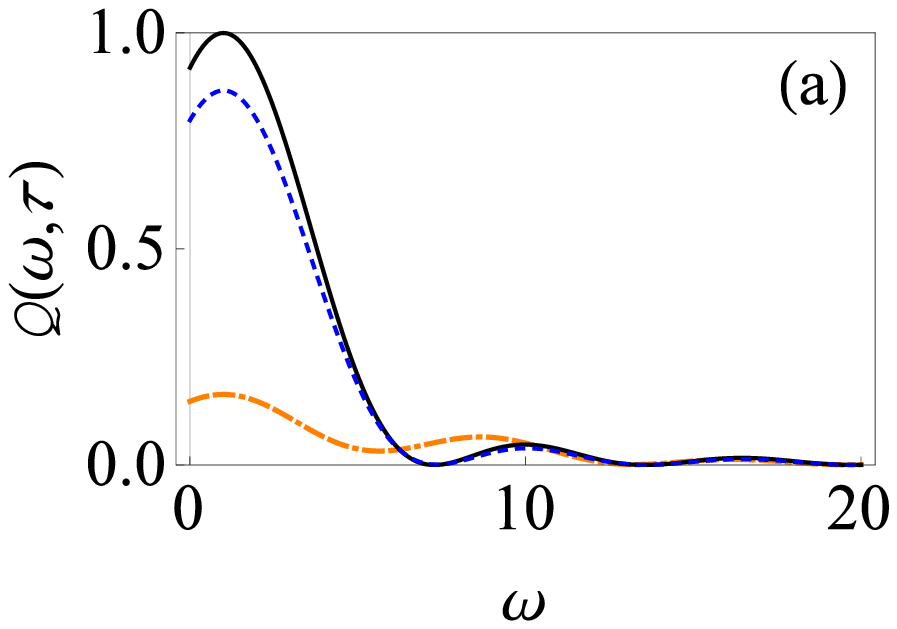}
	\quad
	\includegraphics[scale = 0.85]{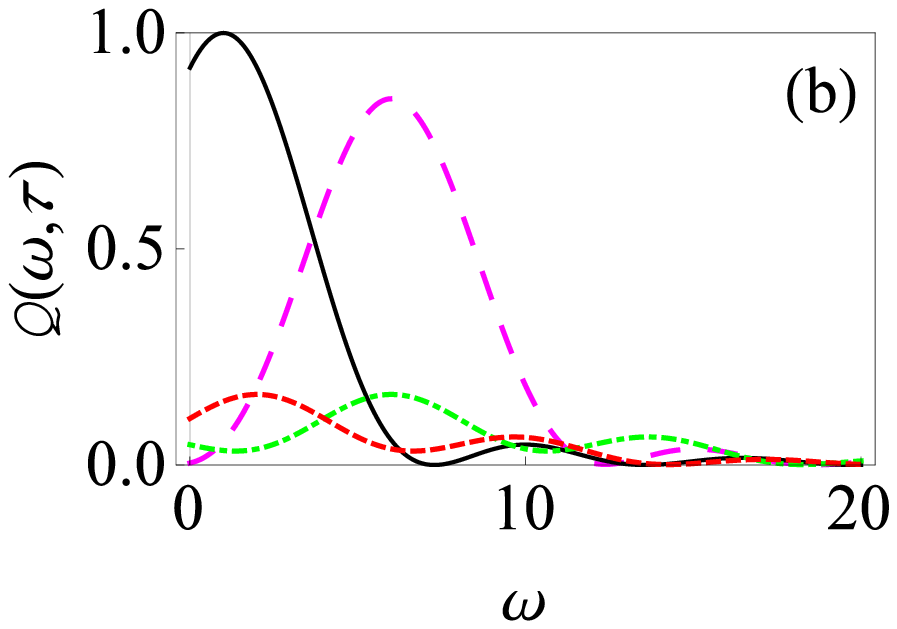}
	
	\centering
	\caption{ \textbf{Filter function for the driven population decay model with the  RWA}. $\textbf{(a)}$  Graph of  $Q(\omega,\tau)$ versus  $\omega$  in the presence of time-dependent driving fields:  $\alpha(t)=\varepsilon_0 t$, $\beta(t)=-(e^{-\chi t}-1)/\chi$ and  $\gamma(t)=0$ with $\chi=0.2$  (dashed, blue curve),  $\alpha(t)=\varepsilon_0 t$, $\beta(t)=\upsilon t$ and  $\gamma(t)=0$  with $\upsilon=5$  (dot-dashed, orange curve), $\alpha(t)=\varepsilon_0 t$, $\beta(t)=0$ and $\gamma(t)=0$ (solid, black curve). We have used $\tau = 1$.   $\textbf{(b)}$  Here $\alpha(t)=\varepsilon_0 t$,  $\beta(t)=\upsilon t$ and $\gamma(t)=\xi t$, with $\upsilon=5$ and $\xi=1$ (dashed, red curve), $\upsilon=1$ and $\xi=5$ (long-dashed, magenta curve), and $\upsilon=5$  and $\xi=5$ (dot-dashed, green curve). The solid black curve shows $\alpha(t)=\varepsilon_0 t$, $\beta(t)=0$ and $\gamma(t)=0$. We have again used $\tau = 1$.}
	\label{Q1ab}
\end{figure}

Let us now consider more complicated driving fields such that we have non-zero values of $\beta(t)$ and $\gamma(t)$. As an example, we consider $\alpha(t)=\varepsilon_0 t$ and $\beta(t)=\upsilon t$, while $\gamma(t) = 0$. This introduces oscillatory fields in the system Hamiltonian [see Eq.~\eqref{EH}], and the filter function now changes as shown in Fig.~\ref{Q1ab}(a). It is clear that adding in the control fields now greatly reduces the filter function (see the dot-dashed orange curve), and is thus expected to lead to a decrease in the effective decay rate. We can also consider what happens if these oscillating control fields are `damped' - this is illustrated by the dashed blue curve. We have checked as that as the fields become more damped, the filter function starts to coincide with the undriven scenario (the solid black curve). Proceeding along these lines, we can also work out the filter function when $\gamma(t)$ is also non-zero, further illustrating the drastic effect of the driving fields on the filter function [see Fig.~\ref{Q1ab}(b)].

\begin{figure}[h!]
	\includegraphics[scale = 0.85]{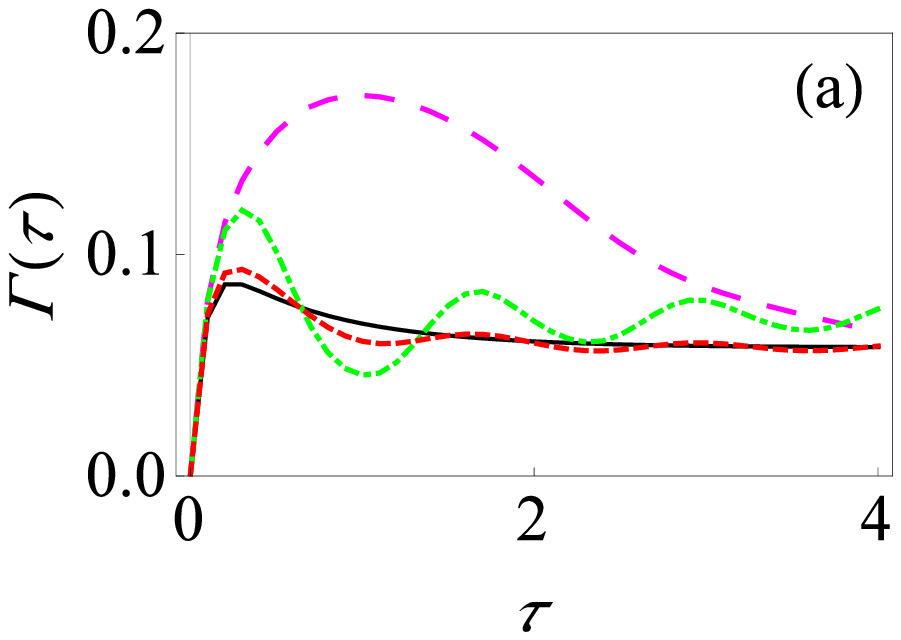}
	\includegraphics[scale = 0.85]{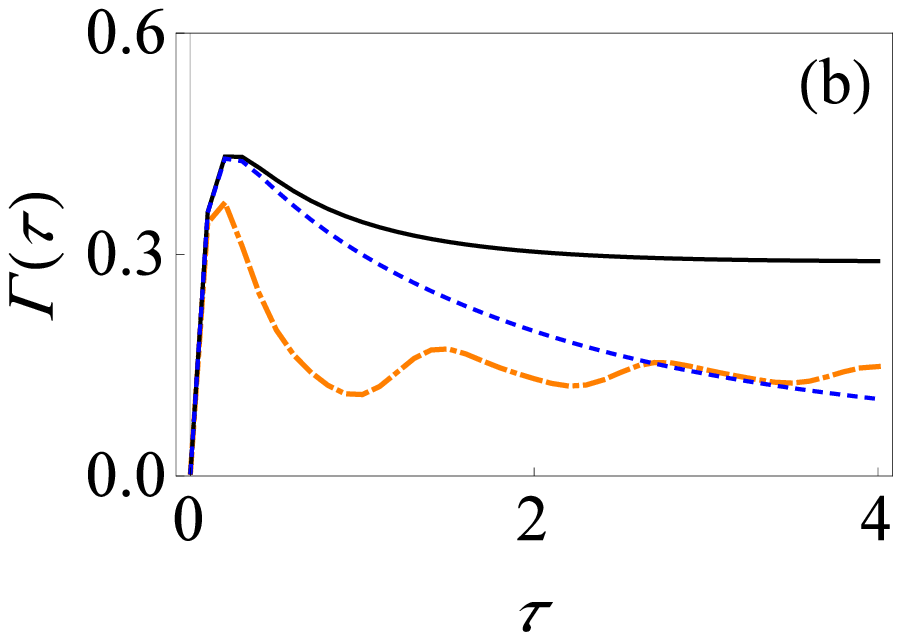}
	
	\centering
	\caption{ \textbf{Effective decay rate for the driven population decay model with the  RWA}. $\textbf{(a)}$  Graph of  $\Gamma(\tau)$ versus  $\tau$  in the presence of time-dependent driving fields:  $\alpha(t)=\varepsilon_0 t +( V_0/\Omega) \sin(\Omega t)$,  $\beta(t)=0$ and $\gamma(t)=0$  with $V_0=1$ and $\Omega=5$ (dashed, red curve), $V_0=5$ and $\Omega=1$ (long-dashed, magenta curve), and $V_0=5$ and $\Omega=5$ (dot-dashed, green curve). The solid black curve shows $\alpha(t)=\varepsilon_0 t$, $\beta(t)=0$ and $\gamma(t)=0$. We remind the reader that we have used dimensionless units with $\hbar = 1$ and we have set $\varepsilon_0 = 1$. Here we have used $G=0.01$ and $\omega_c=10$. $\textbf{(b)}$  We now have $\alpha(t)=\varepsilon_0 t$, $\beta(t)=-(e^{-\chi t}-1)/\chi$ and  $\gamma(t)=0$ with $\chi=0.2$  (dashed, blue curve);    $\alpha(t)=\varepsilon_0 t$, $\beta(t)=\upsilon t$ and  $\gamma(t)=0$  with $\upsilon=5$  (dot-dashed, orange curve); $\alpha(t)=\varepsilon_0 t$, $\beta(t)=0$ and $\gamma(t)=0$ (solid, black curve). We now have $G = 0.05$ and $\omega_c = 10$.} 
	\label{ab}
\end{figure}

Having illustrated the effect of the driving fields on the filter function, we now demonstrate how this translates to a change in the effective decay rate and thereby the quantum Zeno and anti-Zeno behavior. The behavior of the effective decay rate $\Gamma(\tau)$ versus measurement interval $\tau$ is shown in Fig.~\ref{ab} for different driving fields. Let us note how the  behavior of $\Gamma(\tau)$ helps us to identify the quantum Zeno and anti-Zeno regimes. If the effective decay rate $\Gamma(\tau)$ decreases by shortening the measurement interval $\tau$, we are in the quantum Zeno regime while if it increases, then we are in the quantum anti-Zeno regime \cite{SegalPRA2007, ThilagamJMP2010, ChaudhryPRA2014zeno, Chaudhryscirep2016}. Moreover, to actually compute the effective decay rate, we need to specify the spectral density of the environment. Throughout this work, we will use an Ohmic spectral density $J(\omega)=G\omega e^{-\omega/\omega_c}$, where $G$ stands for the dimensionless coupling strength between the system and its environment, and $\omega_c$ is the cutoff frequency. As we have discussed before, for weak system-environment coupling, the effective decay rate is the overlap  integral of the  spectral density  of the environment $J(\omega)$ and  the generalized filter function $Q(\omega,t)$. If the peak value of the filter function is near the cutoff frequency of spectrum of environment then there will be a significant overlap between $J(\omega)$ and $Q(\omega, t)$, giving an enhanced decay rate. On the other hand, if the peak value of generalized filter function is well beyond the $\omega_c$, then that minimizes the overlap between $J(\omega)$ and $Q(\omega, t)$ ,  leading to  a reduced decay rate.   The dynamically modified filter function in the presence of driving fields [see Figs.~\ref{Qab} and \ref{Q1ab}] affects the overlap of $J(\omega)$ with $Q(\omega,t)$, and thus can either accelerate or inhibit the decay rate as compared to the undriven scenario. In particular, it is clear from Fig.~\ref{ab}(a) that for the simple population decay case with no driving fields (the solid black curve), there is a single local crossover between quantum Zeno and anti-Zeno regime, meaning that, in short time regime, effective decay rate $\Gamma(\tau)$ decreases  by decreasing the measurement interval $\tau$ while for large measurement interval, it increases by decreasing the measurement interval. In the presence of driving fields, not only is the effective decay rate greatly affected (see the long-dashed magenta curve), but also there are multiple Zeno and anti-Zeno regimes (see the dashed red curve and the dot-dashed green curve). For long-dashed magenta curve, the peak value of the filter function is at approximately $\omega=5.6$ [see Fig.~\ref{Qab}(b)] which is more near to the peak of the spectrum of the environment as compared to the solid black curve (for which the peak is at $\omega=1$). As a result, this gives maximum overlap of $Q(\omega,\tau)$ with $J(\omega) $ for the magenta curve, which consequently enhances the effective decay rate compared to the undriven case. For the dot-dashed green curve, we observed previously that a large driving field frequency means that the peak of the filter function keeps changing as the measurement interval changes, and this leads to multiple Zeno and anti-Zeno regimes. However, at short measurement intervals, all the curves agree. This is expected, since, as we have seen before, with very fast measurements, the filter function becomes the same, leading to the same decay rate. We have also looked at what happens with $\beta(t) \neq 0$. Having seen how the filter function is influenced by such driving fields in Fig.~\ref{Q1ab}(a), we illustrate what happens to the corresponding decay rate in Fig.~\ref{ab}(b) for the same $\beta(t)$ as used in Fig.~\ref{Q1ab}(a), where $\beta(t)$ is a damped function  for the dashed blue curve, and it is a linear function of $t$ for the dot-dashed orange curve. Since a linear function of $t$ in the presence of non-zero $\varepsilon_0$ leads to a reduction in the peak value of the filter function [see Fig.~\ref{Q1ab}(a)], the overlap  between $Q(\omega,\tau)$ and $J(\omega)$ reduces for $\omega_c=10$, leading to a reduction in the effective decay rate as compared to the solid black curve. On the other hand, if $\beta(t)$ is a damped function, effects of the oscillating fields are suppressed, meaning that the effective decay rate is increased for the dashed blue curve as compared to dot-dashed orange curve. Once again, it is clear that the driving fields greatly influence the decay rate both quantitatively and qualitatively.

We next discuss the effect of the non-rotating terms of the system-environment interaction on the dynamics of the central system in the presence of a driving field. The total Hamiltonian is now given by 
\begin{equation}
H(t) = \frac{\varepsilon_0}{2}\sigma_z +H_c(t)+ \sum_k \omega_k b_k^\dagger b_k +\sigma_x \sum_k (g_k^*b_k + g_k b_k^\dagger).
\end{equation}\\
Notice the different form of the system-environment coupling as compared to before - the system-environment Hamiltonian now contains the `non-rotating' terms $\sigma_+ b_k^\dagger$ and $\sigma_- b_k$. To calculate filter function $Q(\omega,t)$ now, we first evaluate the environment correlation function. With $F=\sigma_x$, we find $G(t)=- e^{-i(\alpha(t)-\gamma(t))} \sin^2[\beta (t)/2]+e^{i(\alpha(t)+\gamma(t))}\cos^2[\beta (t)/2]$ and $\bar{G}(t-t')=- e^{i(\alpha(t-t')-\gamma(t-t'))} \sin^2[\beta(t-t')/2]+e^{-i(\alpha(t-t')+\gamma(t-t'))}\cos^2[\beta(t-t')/2]$. Also, $f(\omega,t) = e^{-i\omega t}$ at zero temperature. Putting all this together,
\begin{align}\label{nonrotating}
Q(\omega,\tau) =\frac{2}{\tau} \int_0^\tau dt \int_0^t dt'\bigr( D_1(t,t')+D_2(t,t')+D_3(t,t')+D_4(t,t')\bigr),
\end{align}\\
where
\begin{align}\label{JQ1}
D_1(t,t')=\cos[\gamma(t)-\gamma(t-t')-\omega t']\cos[\alpha(t)]\cos[\beta(t)]\cos[\alpha(t-t')]\cos[\beta(t-t')],
\end{align}
\begin{equation}\label{JQ2}
D_2(t,t')=-\sin[\gamma(t)-\gamma(t-t')-\omega t']\cos[\alpha(t-t')]\cos[\beta(t-t')]\sin[\alpha(t)],
\end{equation}
\begin{equation}\label{JQ3}
D_3(t,t')=\sin[\gamma(t)-\gamma(t-t')-\omega t']\cos[\alpha(t)]\cos[\beta(t)]\sin[\alpha(t-t')],
\end{equation}
\begin{equation}\label{JQ4}
D_4(t,t')=\cos[\gamma(t)-\gamma(t-t')-\omega t']\sin[\alpha(t)]\sin[\alpha(t-t')].
\end{equation}
Compared with Eq.~\eqref{rotating}, we can see that the two filter functions agree for $\beta(t)=0$. This means that in the absence of driving fields (where $\alpha(t) = \varepsilon_0 t$ and $\beta(t) = 0$), the counter-rotating terms do not affect the decay rate. However, in the presence of driving fields with $\beta(t) \neq 0$, the counter-rotating terms become important, even in the weak system-environment coupling regime we are dealing with. The influence of the non-rotating terms is shown in Fig.~\ref{Rab}, where the behavior of the filter function as a function of the frequency $\omega$ is shown in Fig.~\ref{Rab}(a) and  the behavior of  the effective decay rate as a function of  measurement interval is illustrated in Fig.~\ref{Rab}(b). Comparing Figs.~\ref{Q1ab}(a) and Fig.~\ref{Rab}(a), it is clear that when there are no driving fields, the filter function does not change since the solid, black curve is the same in both figures. However, as shown by the dashed blue and dot-dashed orange curves, in the presence of driving fields with $\beta(t) \neq 0$, the filter function does change. This correspondingly changes the effective decay rate by modifying the overlap of $J(\omega)$ with $Q(\omega, t)$ as can be seen by comparing Figs.~\ref{ab}(b) and \ref{Rab}(b). The counter-rotating terms help to enhance the peak value of the  filter function, leading to an increase in the effective decay rate.

\begin{figure}[t]
	\includegraphics[scale = 0.85]{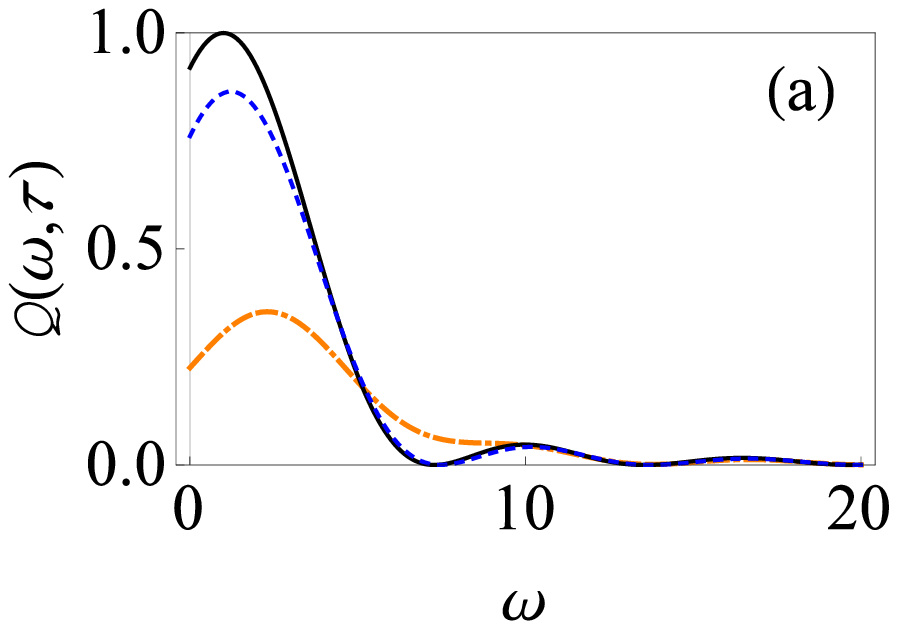}
	\quad
	\includegraphics[scale = 0.85]{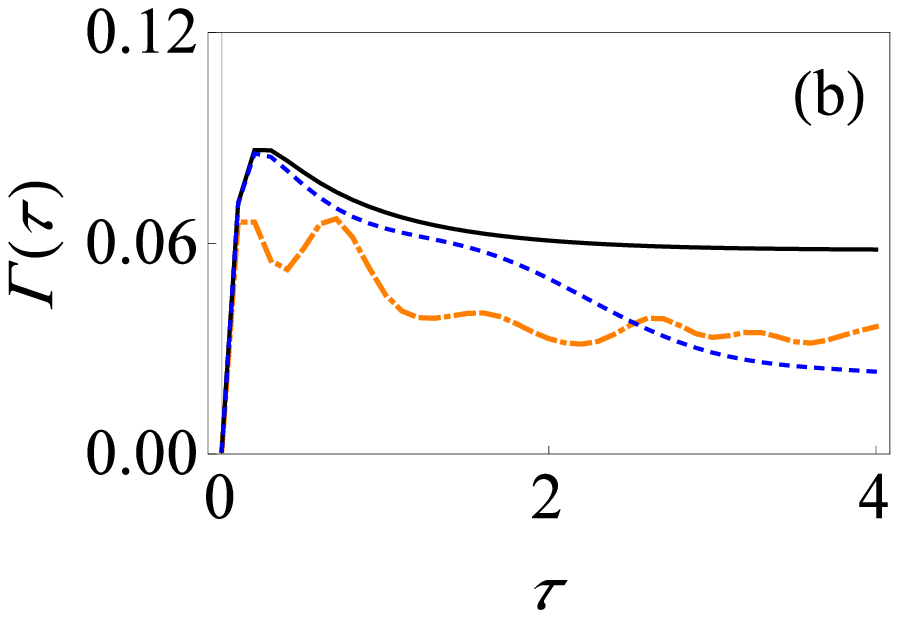}
	
	\centering
	\caption{\textbf{Filter function and effective decay rate for the driven population decay model without the RWA}. $\textbf{(a)}$  Graph of $Q(\omega,\tau)$ versus  $\omega$  in the presence of time-dependent driving fields: $\alpha(t)=\varepsilon_0 t$, $\beta(t)=-(e^{-\chi t}-1)/\chi$ and  $\gamma(t)=0$ with $\chi=0.2$  (small-dashed, blue curve);  $\alpha(t)=\varepsilon_0 t$, $\beta(t)=\upsilon t$ and  $\gamma(t)=0$  with $\upsilon=5$  (dot-dashed-dashed, orange curve); $\alpha(t)=\varepsilon_0 t$, $\beta(t)=0$ and $\gamma(t)=0$ (solid, black curve). Here we have used $\tau=1$. $\textbf{(b)}$   Similar to $\textbf{(a)}$, except that now we have plotted $\Gamma(\tau)$ versus $\tau$ with $G=0.01$ and $\omega_c=10$.}
	\label{Rab}
\end{figure}

\subsection*{Application to the driven dephasing model in weak coupling regime}	
We now consider the pure dephasing model given by the system-environment Hamiltonian
\begin{equation}\label{ddr}
H = \frac{\varepsilon_0}{2} \sigma_z+ \sum_k \omega_k b_k^\dagger b_k +\sigma_z \sum_k (g_k^*b_k + g_k b_k^\dagger).
\end{equation}
Notice the different form of the system-environment interaction. With this model, the populations of the system energy eigenstates cannot change - only the off-diagonal coherences can change, which is why this is referred to as a pure dephasing model\cite{ChaudhryPRA2014zeno}. The initial state usually considered in this model is $\ket{\psi}=\dfrac{1}{\sqrt{2}}(\ket{e}+\ket{g})$, with $\bra{e}g \rangle=0$. However, with the formalism we have developed, the initial state we considered was $\ket{e}$. To use our formalism, we consequently perform a unitary operation $U_R = e^{i\pi \sigma_y/4}$. The initial state then again becomes $\ket{e}$, while the Hamiltonian is transformed to 
\begin{equation}\label{rotated}
H^{(R)} = U^{(R)}HU^{(R)\dagger} = -\frac{\varepsilon_0}{2} \sigma_x+ \sum_k \omega_k b_k^\dagger b_k -\sigma_x \sum_k (g_k^*b_k + + g_k b_k^\dagger).
\end{equation}
To find the filter function now, we look at Eq.~\eqref{EH} and find that $\alpha(t) = \pi/2$, $\beta(t) = \varepsilon_0 t$ and $\gamma(t) = -\pi/2$ gives the same Hamiltonian as Eq.~\eqref{rotated}. Then, using our developed formalism, we find that $G(t)=1$ and $\bar{G}(t-t')=1$. Consequently, assuming that the environment is at zero temperature, we get $Q(\omega,\tau)=\frac{2}{\tau}\frac{1-\cos(\omega \tau)}{\omega^2}$, which agrees with the filter function obtained using the exact solution \cite{Chaudhryscirep2016}. Next, we add in the effect of the driving fields. To this end, we look at more complicated time-dependent functions $\alpha(t)$, $\beta(t)$, and $\gamma(t)$. We write the corresponding system-environment Hamiltonian as 
\begin{equation}\label{PN}
H(t) = -\frac{\varepsilon_0}{2} \sigma_x+ H_c(t)+\sum_k \omega_k b_k^\dagger b_k -\sigma_x \sum_k (g_k^*b_k + + g_k b_k^\dagger). 
\end{equation}
To take into account the additional control fields given by $H_c(t)$, we write $\alpha(t) = \pi/2+ \widetilde{\alpha}(t)$ and $\gamma(t)=-\pi/2+\widetilde{\gamma}(t)$, while $\beta(t)$ remains $\varepsilon_0 t$. Simple calculations then lead to the filter function  
\begin{align}
Q(\omega,\tau) =\frac{2}{\tau} \int_0^\tau dt \int_0^t dt'\bigr( D_1(t,t')+D_2(t,t')+D_3(t,t')+D_4(t,t')\bigr),
\end{align}\\
where now
\begin{align}
D_1(t,t')=\cos[\widetilde{\gamma}(t)-\widetilde{\gamma}(t-t')-\omega t']\sin[\widetilde{\alpha}(t)]\cos[\beta(t)]\sin[\widetilde{\alpha}(t-t')]\cos[\beta(t-t')],
\end{align}
\begin{equation}
D_2(t,t')=\sin[\widetilde{\gamma}(t)-\widetilde{\gamma}(t-t')-\omega t']\sin[\widetilde{\alpha}(t-t')]\cos[\beta(t-t')]\cos[\widetilde{\alpha}(t)],
\end{equation}
\begin{equation}
D_3(t,t')=-\sin[\widetilde{\gamma}(t)-\widetilde{\gamma}(t-t')-\omega t']\sin[\widetilde{\alpha}(t)]\cos[\beta(t)]\cos[\widetilde{\alpha}(t-t')],
\end{equation}
\begin{equation}
D_4(t,t')=\cos[\widetilde{\gamma}(t)-\widetilde{\gamma}(t-t')-\omega t']\cos[\widetilde{\alpha}(t)]\cos[\widetilde{\alpha}(t-t')].
\end{equation}
Using these expressions, we have plotted the filter function (for $\tau = 1$) in Fig.~\ref{dab}(a) for different control fields. Once again, it is clear that the driving fields greatly influence the filter function in general. For instance, with a sinusoidal driving field ($\widetilde{\alpha}(t)=(V_0/\Omega) \sin(\Omega t)$, $\beta(t)=\varepsilon_0 t$, and  $\widetilde{\gamma}(t)=0$), the filter function is very different as compared with the undriven case (compare the solid black curve with the dot-dashed green and long-dashed magenta curves), with the difference becoming smaller as the driving field strength is reduced (see the dashed red curve). Consequently, the behavior of effective decay rate and the corresponding quantum Zeno and anti-Zeno phenomena is expected to be greatly modified due to the different overlap of the filter function with the spectrum of the environment. That this is indeed the case can be seen in Fig.~\ref{dab}(b). We see that there is a single peak in the case of the undriven pure dephasing model. On the other hand, due to the application of external fields, the effective decay rate increases with increasing $V_0$  and the effective decay rate shows multiple Zeno and anti-Zeno regimes for fast oscillating external fields.

\begin{figure}[t]\includegraphics[scale = 0.85]{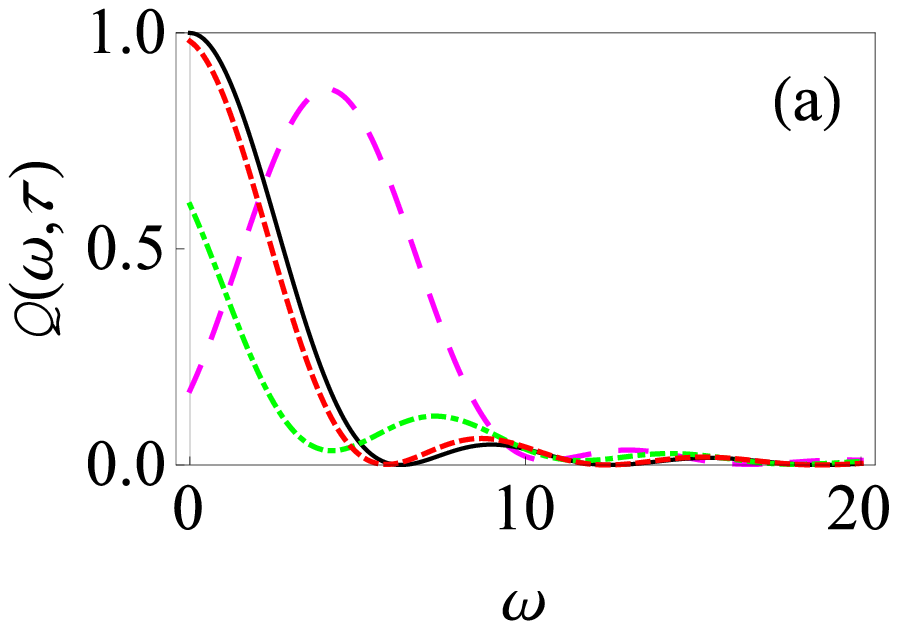}
	\quad
	\includegraphics[scale = 0.85]{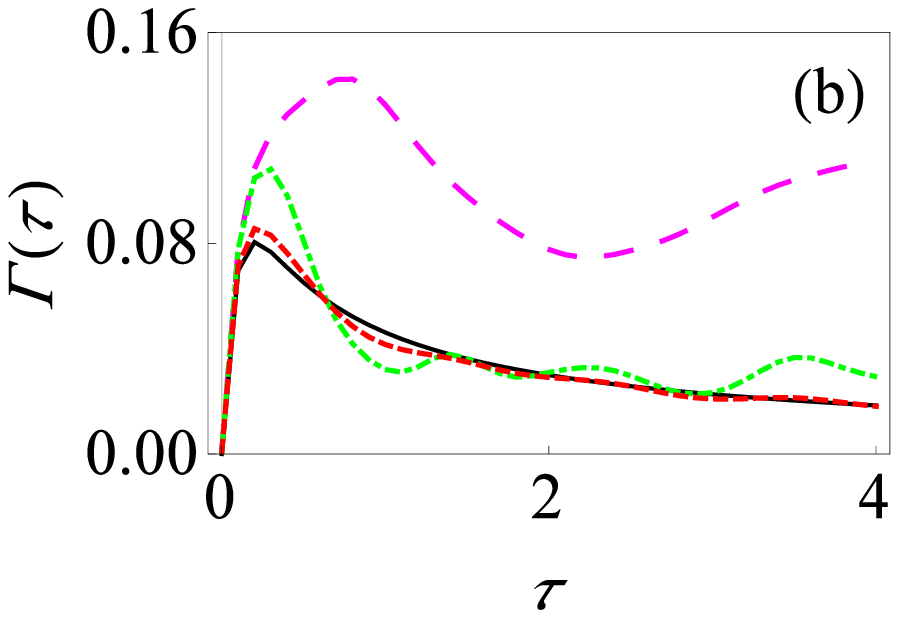}
	\centering
	\caption{ \textbf{Filter function and effective decay rate for the driven dephasing model}. $\textbf{(a)}$  Graph of $Q(\omega,\tau)$ versus  $\omega$  in the presence of time-dependent driving fields: $\widetilde{\alpha}(t)=( V_0/\Omega) \sin(\Omega t)$, $\beta(t)=\varepsilon_0 t$ and  $\widetilde{\gamma}(t)=0$ with $V_0=1$ and $\Omega=5$ (dashed, red curve), $V_0=5$ and $\Omega=1$ (large-dashed, magenta curve), and $V_0=5$ and $\Omega=5$ (dot-dashed, green curve). The solid black curve is with $\widetilde{\alpha}(t)=0$, $\beta(t)=\varepsilon_0 t$, and $\widetilde{\gamma}(t)=0$. Again, we are using dimensionless units with $\hbar = 1$ and we have set $\varepsilon_0 = 1$. Here we are have used $\tau=1$.  $\textbf{(b)}$ Similar to $\textbf{(a)}$, except that now we have plotted $\Gamma(\tau)$ versus $\tau$ with $G=0.01$ and $\omega_c=10$.}
	\label{dab}
\end{figure}


\subsection*{Application to the driven large spin-boson model in weak coupling regime}

We now briefly show how we can extend our formalism to more general systems in which a large spin greater than spin-1/2
is coupled to a harmonic oscillator environment. Such a model can describe, for instance, a collection of $N_S$ two-level systems coupled to a common environment\cite{ChaudhryPRA2014zeno,VorrathPRL2005,KurizkiPRL2011}. We first define $J_k=\dfrac{1}{2} \sum_i\sigma_{k}^{(i)}$, where $J_k$ ($k = x, y, z$) are the large spin operators. As a concrete example, we consider the driven population decay model given by the system-environment Hamiltonian 
\begin{equation}\label{NTWA}
H(t) = {\varepsilon_0}J_z +H_c(t) +\sum_k \omega_k b_k^\dagger b_k + 2 J_x \sum_k (g_k^*b_k  + g_k b_k^\dagger),
\end{equation}
where $\varepsilon_0$ is the  energy level spacing for each spin-1/2 particle, and  $H_c(t)$ is the control field Hamiltonian. Analogous to what we did for the single spin-1/2 case, we consider the free system unitary time evolution operator to be $U_S(t) = e^{-i\alpha(t) J_z} e^{-i\beta(t) J_y} e^{-i\gamma(t)J_z}$. We take the initial state to be $\ket{j}$ with $J_z \ket{j} = j\ket{j}$ and $j = N_S/2$. Performing the calculation for the filter function using our formalism, we find that the filter function is exactly the same as for the single spin-1/2 case [see Eq.~\eqref{nonrotating}] except for an additional multiplicative factor of $N_S$ (see the Methods section for details). That is, the effective decay rate is now enhanced by a factor of $N_S$, reminiscent of the superradiance effect \cite{Dicke1954}. A similar calculation shows that if the pure dephasing model is extended to the large spin case in an analogous manner, the effective rate is again enhanced by a factor of $N_S$.

\subsection*{Application to the driven spin-boson model in the strong system-environment coupling regime}
Finally, let us extend our treatment of the driven population decay model to the strong system-environment case. We start from the system-environment Hamiltonian $H(t) = H_S(t) + H_B + H_{SB}$ with $H_{S}(t)=\dfrac{\varepsilon(t)}{2}\sigma_z+\dfrac{\Delta}{2}\sigma_x$, $H_{B}=\sum_k\omega_kb_k^\dagger b_k$, and $H_{SB}=\sigma_z\sum_k(g^*_kb_k+g_kb^\dagger_k)$. The driving fields are contained in $\varepsilon(t)$. Now, if the system and the environment are interacting strongly, we cannot treat the interaction term between system and its environment perturbatively. Instead, we can consider performing a polaron transformation \cite{Vorraththesis,ChinPRL2011,LeeJCP2012,GuzikJPCL2015,Chaudhryscirep2017a}, which transforms our Hamiltonian to (see the supplementary material)
\begin{equation}\label{PT}
H^{(P)}(t)=e^{\chi\sigma_z/2}H(t)e^{-\chi\sigma_z/2} = \dfrac{\varepsilon(t)}{2}\sigma_z+\sum_k\omega_kb^\dagger_kb_k+\dfrac{\Delta}{2}(\sigma_+Y+\sigma_-Y^\dagger), 
\end{equation}\\
where $P$ denotes the so-called polaron frame, $\chi=\sum_k[\dfrac{2g_k}{\omega_k}b_k^\dagger-\dfrac{2g^*_k}{\omega_k}b_k]$, and $Y = e^\chi$. 

We see that in the polaron frame, the form of the system-environment interaction is different. Now, if the tunneling parameter $\Delta$ is small, we can apply time-dependent perturbation theory, treating the system-environment coupling in the polaron frame perturbatively. As before, at the initial time, we prepare our system in the excited state $\ket{e}$, and then perform  projective measurements on the system with time interval $\tau$ to check whether the system is still present in excited state $\ket{e}$ or not. It is also important to note that initial state of system and environment in the untransformed frame cannot be written in the simple usual product form $\rho(0)=\ket{e}\bra{e}\otimes e^{-\beta H_B}/Z_B$, with  $\rho_B=e^{-\beta H_B}/Z_B$ and  $Z_B=\tr_B[e^{-\beta H_B}]$, since the system and the environment are strongly interacting in that frame and consequently, the initial system-environment correlations are significant \cite{PollakPRE2008,ChaudhryPRA2013a,ChaudhryPRA2013b,ChaudhryEJPD2019b,ChaudhryPRA2020}.  However, since the system and its environment are effectively weakly interacting in the polaron frame, the initial state in the polaron frame can be taken as a simple product state. The rest of the calculation, performed in the polaron frame, proceeds in a similar way as the weak coupling case using perturbation theory. We eventually arrive at  (see the supplementary Material)
\begin{equation}\label{PRI}
\Gamma(\tau)=\dfrac{\Delta^2}{2\tau}\int_0^\tau dt\int_0^t dt'e^{-\Phi_R(t')}\cos[\zeta(t)-\zeta(t-t')-\Phi_I(t')],
\end{equation}
where $\Phi_I(t)=\int_0^\infty d\omega J(\omega)\dfrac{\sin(\omega t)}{\omega^2}$, $\Phi_R(t)=\int_0^\infty d\omega J(\omega)\dfrac{1-\cos(\omega t)}{\omega^2}\coth(\dfrac{\beta\omega}{2})$, and $\zeta(t) = \int_0^t \varepsilon(t')\, dt'$.
Assuming, as before, an Ohmic spectral density of the form $J(\omega)=G\omega e^{-\omega/\omega_c}$,  $\Phi_I(t)$ and  $\Phi_R(t)$ are found to be (at zero temperature) $\Phi_I(t)=G\tan^{-1}(\omega_ct)$ and $\Phi_R(t)=\frac{G}{2}\ln(1+\omega_c^2t^2)$.

\begin{figure}[t]
	\includegraphics[scale = 0.85]{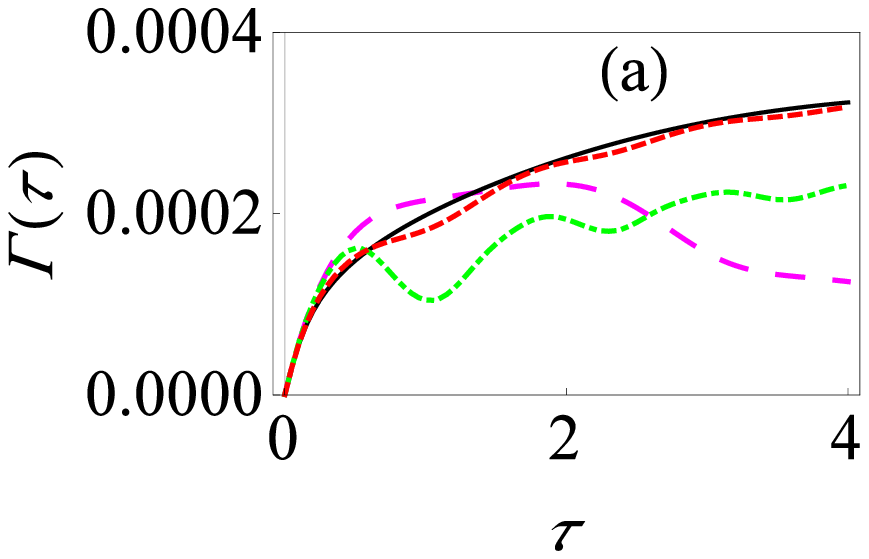}
	\quad
	\includegraphics[scale = 0.85]{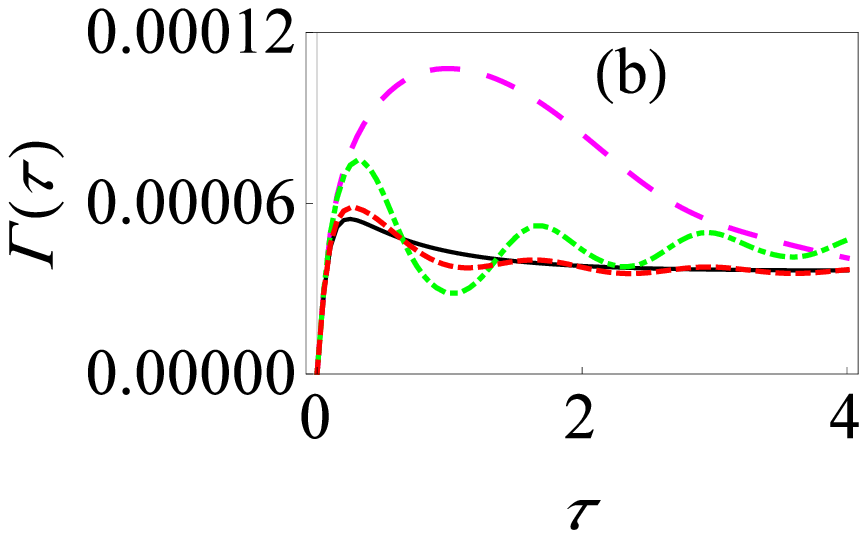}
	\centering
	\caption{ \textbf{Effective decay rate for the strongly coupled driven  spin-boson model}. $\textbf{(a)}$ Graph of  $\Gamma(\tau)$ versus  $\tau$  in the presence of the time-dependent driving fields:  $\alpha(t)=\varepsilon_0 t +( V_0/\Omega) \sin(\Omega t)$,  $\beta(t)=0$ and $\gamma(t)=0$  with $V_0=1$ and $\Omega=5$ (dashed, red curve), $V_0=5$ and $\Omega=1$ (large-dashed, magenta curve), and $V_0=5$ and $\Omega=5$ (dot-dashed, green curve). The solid black curve shows $\alpha(t)=\varepsilon_0 t$, $\beta(t)=0$ and $\gamma(t)=0$. Here we have used $\Delta=0.05$, $\omega_c=10$ and $G=1$.  $\textbf{(b)}$ Same as $\textbf{(a)}$, except that now $G=2$.}
	\label{Pab}
\end{figure}

We now have everything we need to calculate the effective decay rate. It should be obvious from our expressions above that for the strong system-environment coupling regime that we are considering here, the effective decay rate $\Gamma(\tau)$ no longer depends on an overlap integral of the generalized filter function with the spectral density of environment. Rather, there is now a non-linear dependence on the spectral density, leading to very different qualitative behavior as compared with the usual weak system-environment coupling regime. For instance, as
	the system-environment coupling strength increases, $e^{-\Phi_R(t)}$ decreases, and thus we anticipate $\Gamma(\tau)$ to decrease. Most importantly for us, we expect that the driving fields have a drastic, non-trivial effect not only the value of the effective decay rate, but also quantum Zeno and anti-Zeno behavior since the integrand in Eq.~\eqref{PRI} obviously depends on the driving fields. To make these claims concrete, let us show the quantitative behavior of effective decay rate $\Gamma(\tau)$. 
It is clear from Fig.~\ref{Pab} that not only the driving fields affect the decay rate very significantly, but also that increasing the system-environment coupling strength $G$ reduces the effective decay rate [compare Figs.~\ref{Pab}(a) and (b)], in contrast with the weak coupling regime. Here again, we observe multiple oscillations in quantum Zeno
to anti-Zeno regimes, as we increase $\Omega$.
 Interestingly, Fig.~\ref{ab}(a) looks very similar with Fig.~\ref{Pab}(b) despite having coupling strength $G$ in different regimes. This similarity can be understood noting that in both cases, the pointer states are the eigenstates of the operator $\sigma_z$, and that in the strong coupling regime, we have a population decay model in the polaron frame (see Eq.~\eqref{PT}). In both cases, the driving fields modulate the energy-level splitting. These similarities leads to the same qualitative form of Fig.~\ref{ab}(a) and Fig.~\ref{Pab}(b).

\section{Discussion}
In this paper, we started off by introducing a general formalism to calculate the effective decay rate of a quantum system subjected to both periodic projective measurements and driving fields, valid for weak system-environment coupling strength. We then applied this formalism to derive a general expression for the decay rate of a driven two-level system. The decay rate is an overlap integral of the spectral density of environment and a filter function expressed using time-dependent Euler angles. These results were illustrated using the population decay model as well as the pure dephasing model. In both cases, the application of the driving fields very significantly changes the form of filter function of the central system, which then modifies (either enhances or minimizes) the effective decay rate and, consequently, the quantum Zeno and anti-Zeno regimes are altered. Interestingly, for the population decay model, the driving fields can lay bare the effect of the counter-rotating terms, even for the weak system-environment coupling regime that we are dealing with. These results were then generalized to large spin systems to show a possible amplification of the decay rate. Finally, we also looked at driven two-level systems strongly coupled to an environment of harmonic oscillators,   where the effective decay rate shows a non-linear
	dependence on the spectral density of the environment. We showed once again that the effective decay rate is modified by the application of driving fields. Our general expressions and insights should of interest in the broad areas of quantum control and quantum state engineering, such as quantum noise sensing, as well as in fundamental studies of the quantum Zeno and anti-Zeno effects. For example, a quantum system can be put into the Zeno regime, thereby protecting it from decoherence, by applying suitable control fields. On the other hand, the decay rate can be enhanced in the anti-Zeno regime via the applied control fields, and this can be useful for cooling the quantum system \cite{Kurizki2015}.

\section{Methods}

\subsection*{Effective decay rate using perturbation theory}

We first discuss how to obtain effective decay rate of system under frequent measurements, extending the treatment in Ref.~{\renewcommand{\citemid}{}\cite[]{Chaudhryscirep2016} to time-dependent Hamiltonians. We first write the system-environment Hamiltonian as $H(t) = H_F(t) + H_{SB}$, where $H_F(t) = H_S(t) + H_B$ is the sum of the free system and environment Hamiltonians and $H_{SB}$ describes the system-environment interaction. Using the standard perturbation approach, we set $U(t)=U_F(t)U_I(t)$, where ${U}_F(t)$ describes the free time evolution of driven system and environment (this may be non-trivial and involve time-ordering due to the possible time dependence of $H_S$), and ${U}_{I}(t)$ is the leftover part that can be expanded perturbatively (assuming the system-environment interaction to be weak). Expanding up to second order in system-environment
coupling, we can write ${U}_{I}(t)=\mathds{1}+{G}_{1}+{G}_{2}$, where $G_1$ and $G_2$ are the first, and second order corrections respectively. Using this expansion, the density matrix at time $\tau$ can be written as 
\begin{align*}\label{NTS6}
{\rho}_{S}(\tau)\approx\tr_{B}\bigr[{U}_{F}(\tau)\bigr({\rho}(0)+{\rho}(0){G}^{\dagger}_{1}+{G}_{1}{\rho}(0)+{\rho}(0){G}^{\dagger}_{2}+{G}_{2}{\rho}(0)+{G}_{1}{\rho}(0){G}^{\dagger}_{1}\bigr){U}^{\dagger}_{F}(\tau)\bigr].
\end{align*}
Perturbation theory tells us that ${G}_{1}=-i\int^{\tau}_{0}dt_{1}\widetilde{H}_{SB}(t_{1})$, and ${G}_{2}=-\int^{\tau}_{0}dt_{1}\int^{t_{1}}_{0}dt_{2}\widetilde{H}_{SB}(t_{1})\widetilde{H}_{SB}(t_{2})$, where $\widetilde{H}_{SB}(t)={U}^{\dagger}_{F}(t){{H}}_{SB} {U}_{F}(t)=\sum_\mu{U}_{S}^{\dagger}(t)F_\mu {U}_{S}(t)\otimes{{U}}^{\dagger}_{B}(t)B_\mu{{U}}_{B}(t)={\widetilde{F}_\mu}(t){\widetilde{B}}_\mu(t)$. Each term can then be simplified to eventually arrive at 
\begin{equation}
\rho_S(\tau) = U_S(\tau) \biggl( \rho_S(0) + \sum_{\mu \nu}\int_0^\tau dt_1 \int_0^{t_1} dt'  C_{\mu \nu}(t') 
[\widetilde{F}_\nu(t_1-t')\rho_S(0),\widetilde{F}_\mu(t_1)] + \text{h.c.} \biggr) U_S^\dagger (\tau).
\label{densitymatrixattau}
\end{equation}
where $C_{\mu \nu}(t') = \tr_B [\frac{e^{-\beta H_B}}{Z_B} \widetilde{B}_\mu(t')\widetilde{B}_\nu(0)]$ are the environment correlation functions, and $\text{h.c.}$ denotes hermitian conjugate. 

Once we have the expression for the system density matrix at time $\tau$, we can evaluate the survival probability of the system in the initial state. Since we are interested to investigate the system evolution due to system-environment interaction only, we first apply the free driven system unitary operator on both sides of Eq.~\eqref{densitymatrixattau} to remove the system evolution due to free driven system Hamiltonian itself. Note that this may not be necessary if $U_S(\tau)$ commutes with $\rho_S(0)$. We thereafter find the probability that the system is still in the initial state $\ket{\psi}$ after a projective measurement given by the projector $\ket{\psi}\bra{\psi}$ to be given by
\begin{align}
s(\tau) = 1-2\text{Re}\,\biggr(
\sum_{\mu \nu}\int_0^\tau dt \int_0^{t} dt'  C_{\mu \nu}(t') 
\tr_S\bigr[P_{\perp}\widetilde{F}_\nu(t-t')\rho_S(0)\widetilde{F}_\mu(t) \bigr]\biggr),
\end{align}\\
where $P_{\perp}$ is the projector onto the subspace orthogonal to $\ket{\psi}\bra{\psi}$. After performing a sequence of $M$  repeated projective measurements, we find the survival probability that the system state is still present in the initial state is $S(M\tau)=[s(\tau)]^M$ if the system-environment correlations are ignored during the evolution.  We can then  find the effective decay rate $\Gamma(\tau)$ by $S(M\tau)=e^{-\Gamma(\tau)M\tau}$ allowing us to write $\Gamma(\tau)=-\ln s(\tau)/\tau$.  In the weak system-environment coupling regime, we can further write
\\\begin{align*}\label{EDRT}
\Gamma(\tau) =\frac{2}{\tau}\text{Re}\biggr(
\sum_{\mu \nu}\int_0^\tau dt \int_0^{t} dt'  C_{\mu \nu}(t') 
\tr_S\bigr[P_{\perp}\widetilde{F}_\nu(t-t')\rho_S(0)\widetilde{F}_\mu(t) \bigr]\biggr).
\end{align*}\\
This can be cast into the form
\begin{equation*}
\Gamma(\tau) = \int_0^\infty \, d\omega \,Q(\omega,\tau) J(\omega),
\end{equation*}\\
with $Q(\omega,\tau)$ given in Eq.~\eqref{Filterfunction}.

\subsection*{Finding the filter function for the driven large-spin population decay model}
In this case, using the standard angular momentum relations $[J_i,J_j]=i\varepsilon_{ijk}J_k$, we evaluate 
$$\widetilde{F}(t) =2 \bigr(J_xc_x(t)+J_yc_y(t)+J_zc_z(t)\bigr),$$
where
\begin{align}
c_x(t)&= \cos[\alpha(t)]\cos[\beta(t)\cos[\gamma(t)]-\sin[\alpha(t)]\sin[\gamma(t)], \notag \\ c_y(t)&=\cos[\alpha(t)]\cos[\beta(t)\sin[\gamma(t)]+\sin[\alpha(t)]\cos[\gamma(t)],  \notag \\ c_z(t) &=-\cos[\alpha(t)]\sin[\beta(t)].
\end{align} 
For $\rho_S(0)=\ket{j,j}\bra{j,j}$ and $P_{\perp}=\sum_{m=1}^{j-1} \ket{j,m}\bra{j,m}$ with $J_z\ket{j,m}=m\ket{j,m}$, we have   $$\tr_S\bigr[P_{\perp}\widetilde{F}(t-t')\rho_S(0)\widetilde{F}(t) \bigr]=\sum_{m=1}^{j-1} \bra{j,m}\widetilde{F}(t-t')\ket{j,j}\bra{j,j}\widetilde{F}(t)\ket{j,m}.$$
We next note that $\bra{j,j}\widetilde{F}(t)\ket{j,m}= \sqrt{2j}\delta_{j-1,m}\bigr(c_x(t)-ic_y(t)\bigr)$. This leads to the generalized filter function 
\begin{equation*}
Q(\omega,t)=(2j)\frac{2}{\tau} \int_0^\tau dt \int_0^t dt'\bigr( D_1(t,t')+D_2(t,t')+D_3(t,t')+D_4(t,t')\bigr), 
\end{equation*} \\
where expressions of $D_1(t,t')$, $D_2(t,t')$, $D_3(t,t')$ and $D_4(t,t')$	are defined in Eqs.~\eqref{JQ1}-\eqref{JQ4}.

\section*{Acknowledgements}

A.~Z.~C. is grateful for support from HEC under grant No 5917/Punjab/NRPU/R\&D/HEC/2016.

\section*{Author contributions statement}

A.~Z.~C. came up with the basic idea behind this work. M.~M.~ carried out the calculations and plotted the graphs. Both authors contributed towards the writing of the manuscript. 

\section*{Additional information}

\textbf{Competing interests:} The authors declare no competing interests. 

\newpage

\section*{Supplemental Material for `The quantum Zeno and anti-Zeno effects with driving fields in the weak and strong coupling regimes'}

\noindent In this supplemental Material, we use same symbols as introduced in our main text. For completeness, we present
the analytical expressions for the Hamiltonian in the polaron frame. We also outline the calculation of the decay rate $\Gamma(\tau)$ for a single two-level system interacting strongly with an environment of harmonic oscillators.

\subsection*{Spin-boson Hamiltonian in polaron frame}
To transform spin-boson Hamiltonian to polaron frame, we need to find
\begin{equation*}\label{PF}
H^{(P)}(t)=e^{\chi\sigma_z/2}H(t)e^{-\chi\sigma_z/2}.
\end{equation*}
We use the Hadamard lemma
\begin{equation}\label{BCH}
e^AOe^{-A}=O+[A,O]+\dfrac{1}{2!}[A,[A,O]]+....
\end{equation}\\
where $A=\chi\sigma_z/2$, with $\chi=\sum_k[\dfrac{2g_k}{\omega_k}b_k^\dagger-\dfrac{2g^*_k}{\omega_k}b_k]$ and $O=H(t)=\dfrac{\varepsilon(t)}{2}\sigma_z+\dfrac{\Delta}{2}\sigma_x+\sum_k\omega_kb_k^\dagger b_k+\sigma_z\sum_k(g^*_kb_k+g_kb^\dagger)$. \\
We find that
\begin{equation*}
e^{\chi\sigma_z/2}\sigma_ze^{-\chi\sigma_z/2}=\sigma_z,
\end{equation*}		
and		
\begin{align*}
e^{\chi\sigma_z/2}\sigma_xe^{-\chi\sigma_z/2}=\sigma_+e^{\chi}+\sigma_-e^{-\chi}.
\end{align*}		
$\sigma_-$ and $\sigma_+$ are the standard spin-1/2 lowering and raising operators. Carrying on further, we find

\begin{equation*}
e^{\chi\sigma_z/2}(\sum_k\omega_kb^\dagger_kb_k)e^{-\chi\sigma_z/2}=\sum_k\omega_kb^\dagger_kb_k-\sigma_z\sum_k(g^*_kb_k+g_kb^\dagger)+\sum_k\dfrac{|g_k|^2}{\omega_k}.
\end{equation*}		
Similarly

\begin{equation*}
e^{\chi\sigma_z/2}\biggr(\sigma_z\sum_k(g^*_kb_k+g_kb^\dagger)\biggr)e^{-\chi\sigma_z/2}=\sigma_z\sum_k(g^*_kb_k+g_kb^\dagger)-2\sum_k\dfrac{|g_k|^2}{\omega_k},
\end{equation*}
The third term of Eq. (\ref{BCH}) in the above expression is a constant number, so higher order commutators are zero. Now putting all these terms back together, the required Hamiltonian in polaron frame takes the following form\\
\begin{equation*}
H^{(P)}(t)=\dfrac{\varepsilon(t)}{2}\sigma_z+\sum_k\omega_kb^\dagger_kb_k+\dfrac{\Delta}{2}(\sigma_+Y+\sigma_-Y^\dagger)-\sum_k\dfrac{|g_k|^2}{\omega_k},
\end{equation*}\\
with $ Y=e^\chi$, and $\sum_k\dfrac{|g_k|^2}{\omega_k}$ is a constant number term that  gives a constant shift in transformed Hamiltonian, and can thus be dropped.

\subsection*{Effective decay rate of  spin-boson model in polaron frame}

Since system-environment coupling is weak in the polaron frame, we can use expression [see the main text] \\ 
$$
\Gamma(\tau) =\frac{2}{\tau}\text{Re}\biggr(
\sum_{\mu \nu}\int_0^\tau dt \int_0^{t} dt'  C_{\mu \nu}(t') 
\tr_S\bigr[P_{\perp}\widetilde{F}_\nu(t-t')\rho_S(0)\widetilde{F}_\mu(t) \bigr]\biggr),$$\\
to calculate effective decay rate . Here $\rho_S(0)=\ket{e}\bra{e}$ and  $P_\perp=\ket{g}\bra{g}$. We identity $F_1=\dfrac{\Delta}{2}\sigma_+$,  $F_2=\dfrac{\Delta}{2}\sigma_-$, $B_1=Y$, $B_2=Y^\dagger$, $\widetilde{F}_1(t)=\dfrac{\Delta}{2}\sigma_+e^{i\zeta(t)}$ and	 $\widetilde{F}_2(t)=\dfrac{\Delta}{2}\sigma_-e^{-i\zeta(t)}$ with $\zeta(t)=\int_0^tdt'\varepsilon(t')$ leading us to

\begin{align}\label{C12}
\Gamma(\tau) =\frac{2}{\tau}\text{Re}\biggr(
\int_0^\tau dt \int_0^{t} dt'  C_{12}(t') e^{i(\zeta(t)-\zeta(t-t'))} \biggr).
\end{align}\\
To get expression of effective decay rate, the environment correlation function $C_{12}(t)$  needs to be worked out. We now show the details how to find $C_{12}$. As we know $C_{12}(t)=\tr_B[\rho_B\widetilde{B}_1(t)B_2]$, with $B1=Y$, $B_2=Y^\dagger$, 	$\widetilde{B}_1(t)=e^{iH_B^{(P)}t}Y e^{-iH_B^{(P)t}}$, $H^{(P)}_B=\sum_k\omega_kb^\dagger_kb_k$, $Y=e^\chi$ and $\chi=\sum_k[\dfrac{2g_k}{\omega_k}b_k^\dagger-\dfrac{2g^*_k}{\omega_k}b_k]$.	Next, we calculate
\begin{equation*}
\widetilde{B}_1(t)=e^{\sum_k\biggr(\frac{g_k}{\omega_k}b^\dagger_ke^{i\omega_kt}-\frac{g^*_k}{\omega_k}b_ke^{-i\omega_kt}\biggr)},
\end{equation*}
\\using the fact $U^\dagger(t)e^AU(t)=e^{U^\dagger(t)AU(t)}$, and then find
\begin{equation*}
\widetilde{B}_1(t)B_2=e^{-i\sum_k\dfrac{|g_k|^2}{\omega_k^2}\sin(\omega_kt)}e^{\sum_k\biggr(\frac{g_k}{\omega_k}b^\dagger_k(e^{i\omega_kt}-1)+\frac{g^*_k}{\omega_k}b_k(e^{-i\omega_kt}-1)\biggr)}.
\end{equation*}		In order to convert double exponential in a single exponential to use useful fact $\tr_B[\rho_Be^C]=e^{\langle C^2\rangle/2}$,  where operator $C$ is a linear combination of annihilation and creation operators; we use the identity $e^Xe^Y=e^{X+Y+\frac{1}{2}[X,Y]}$. Fortunately in this case, first commutator is a constant number, so higher order commutators are zero. Finally we have

\begin{equation*}
C_{12}(t)=e^{-i\Phi_I(t)}e^{-\Phi_R(t)}.
\end{equation*}
where $\Phi_I(t)$	and $\Phi_R(t)$	have been defined in the main text. Carrying on further,we have
\begin{equation}\label{PRI}
\Gamma(\tau)=\dfrac{\Delta^2}{2\tau}\int_0^\tau dt\int_0^t dt'e^{-\Phi_R(t')}\cos[\zeta(t)-\zeta(t-t')-\Phi_I(t')].
\end{equation}


\begin{thebibliography}{10}
\expandafter\ifx\csname url\endcsname\relax
  \def\url#1{\texttt{#1}}\fi
\expandafter\ifx\csname urlprefix\endcsname\relax\def\urlprefix{URL }\fi
\providecommand{\bibinfo}[2]{#2}
\providecommand{\eprint}[2][]{\url{#2}}

\bibitem{Sudarshan1977}
\bibinfo{title}{The {Z}eno's paradox in quantum theory}.
\newblock \emph{\bibinfo{journal}{J. Math. Phys. (N. Y.)}}
  \textbf{\bibinfo{volume}{18}}, \bibinfo{pages}{756} (\bibinfo{year}{1977}).
\newblock \urlprefix\url{http://dx.doi.org/10.1063/1.523304}.

\bibitem{FacchiPhysLettA2000}
\bibinfo{author}{Facchi, P.}, \bibinfo{author}{Gorini, V.},
  \bibinfo{author}{Marmo, G.}, \bibinfo{author}{Pascazio, S.} \&
  \bibinfo{author}{Sudarshan, E.}
\newblock \bibinfo{title}{Quantum {Z}eno dynamics}.
\newblock \emph{\bibinfo{journal}{Phys. Lett. A}}
  \textbf{\bibinfo{volume}{275}}, \bibinfo{pages}{12} (\bibinfo{year}{2000}).
\newblock \urlprefix\url{http://dx.doi.org/10.1016/ S0375-9601(00)00566-1}.

\bibitem{FacchiPRL2002}
\bibinfo{author}{Facchi, P.} \& \bibinfo{author}{Pascazio, S.}
\newblock \bibinfo{title}{Quantum {Z}eno subspaces}.
\newblock \emph{\bibinfo{journal}{Phys. Rev. Lett.}}
  \textbf{\bibinfo{volume}{89}}, \bibinfo{pages}{080401}
  (\bibinfo{year}{2002}).
\newblock \urlprefix\url{http://dx.doi.org/10.1103/PhysRevLett.89.080401}.

\bibitem{FacchiJPA2008}
\bibinfo{author}{Facchi, P.} \& \bibinfo{author}{Pascazio, S.}
\newblock \bibinfo{title}{Quantum {Z}eno dynamics: mathematical and physical
  aspects}.
\newblock \emph{\bibinfo{journal}{J. Phys. A: Math. Theor.}}
  \textbf{\bibinfo{volume}{41}}, \bibinfo{pages}{493001}
  (\bibinfo{year}{2008}).
\newblock \urlprefix\url{http://dx.doi.org/10.1088/1751-8113/41/49/493001}.

\bibitem{WangPRA2008}
\bibinfo{author}{Wang, X.-B.}, \bibinfo{author}{You, J.~Q.} \&
  \bibinfo{author}{Nori, F.}
\newblock \bibinfo{title}{Quantum entanglement via two-qubit quantum {Z}eno
  dynamics}.
\newblock \emph{\bibinfo{journal}{Phys. Rev. A}} \textbf{\bibinfo{volume}{77}},
  \bibinfo{pages}{062339} (\bibinfo{year}{2008}).
\newblock \urlprefix\url{http://dx.doi.org/10.1103/PhysRevA.77.062339}.

\bibitem{ManiscalcoPRL2008}
\bibinfo{author}{Maniscalco, S.}, \bibinfo{author}{Francica, F.},
  \bibinfo{author}{Zaffino, R.~L.}, \bibinfo{author}{Lo~Gullo, N.} \&
  \bibinfo{author}{Plastina, F.}
\newblock \bibinfo{title}{Protecting entanglement via the quantum {Z}eno
  effect}.
\newblock \emph{\bibinfo{journal}{Phys. Rev. Lett.}}
  \textbf{\bibinfo{volume}{100}}, \bibinfo{pages}{090503}
  (\bibinfo{year}{2008}).
\newblock \urlprefix\url{http://dx.doi.org/10.1103/PhysRevLett.100.090503}.

\bibitem{FacchiJPA2010}
\bibinfo{author}{Facchi, P.} \& \bibinfo{author}{Ligab\`{o}, M.}
\newblock \bibinfo{title}{Quantum {Z}eno effect and dynamics}.
\newblock \emph{\bibinfo{journal}{J. Phys. A: Math. Theor.}}
  \textbf{\bibinfo{volume}{51}}, \bibinfo{pages}{022103}
  (\bibinfo{year}{2010}).
\newblock \urlprefix\url{http://dx.doi.org/10.1063/1.3290971}.

\bibitem{MilitelloPRA2011}
\bibinfo{author}{Militello, B.}, \bibinfo{author}{Scala, M.} \&
  \bibinfo{author}{Messina, A.}
\newblock \bibinfo{title}{Quantum {Z}eno subspaces induced by temperature}.
\newblock \emph{\bibinfo{journal}{Phys. Rev. A}} \textbf{\bibinfo{volume}{84}},
  \bibinfo{pages}{022106} (\bibinfo{year}{2011}).
\newblock \urlprefix\url{http://dx.doi.org/10.1103/PhysRevA.84.022106}.

\bibitem{RaimondPRA2012}
\bibinfo{author}{Raimond, J.~M.} \emph{et~al.}
\newblock \bibinfo{title}{Quantum {Z}eno dynamics of a field in a cavity}.
\newblock \emph{\bibinfo{journal}{Phys. Rev. A}} \textbf{\bibinfo{volume}{86}},
  \bibinfo{pages}{032120} (\bibinfo{year}{2012}).
\newblock \urlprefix\url{http://dx.doi.org/10.1103/PhysRevA.86.032120}.

\bibitem{SmerziPRL2012}
\bibinfo{author}{Smerzi, A.}
\newblock \bibinfo{title}{Zeno dynamics, indistinguishability of state, and
  entanglement}.
\newblock \emph{\bibinfo{journal}{Phys. Rev. Lett.}}
  \textbf{\bibinfo{volume}{109}}, \bibinfo{pages}{150410}
  (\bibinfo{year}{2012}).
\newblock \urlprefix\url{http://dx.doi.org/10.1103/PhysRevLett.109.150410}.

\bibitem{WangPRL2013}
\bibinfo{author}{Wang, S.-C.}, \bibinfo{author}{Li, Y.}, \bibinfo{author}{Wang,
  X.-B.} \& \bibinfo{author}{Kwek, L.~C.}
\newblock \bibinfo{title}{Operator quantum {Z}eno effect: Protecting quantum
  information with noisy two-qubit interactions}.
\newblock \emph{\bibinfo{journal}{Phys. Rev. Lett.}}
  \textbf{\bibinfo{volume}{110}}, \bibinfo{pages}{100505}
  (\bibinfo{year}{2013}).
\newblock \urlprefix\url{http://dx.doi.org/10.1103/PhysRevLett.110.100505}.

\bibitem{McCuskerPRL2013}
\bibinfo{author}{McCusker, K.~T.}, \bibinfo{author}{Huang, Y.-P.},
  \bibinfo{author}{Kowligy, A.~S.} \& \bibinfo{author}{Kumar, P.}
\newblock \bibinfo{title}{Experimental demonstration of interaction-free
  all-optical switching via the quantum {Z}eno effect}.
\newblock \emph{\bibinfo{journal}{Phys. Rev. Lett.}}
  \textbf{\bibinfo{volume}{110}}, \bibinfo{pages}{240403}
  (\bibinfo{year}{2013}).
\newblock \urlprefix\url{http://dx.doi.org/10.1103/PhysRevLett.110.240403}.

\bibitem{StannigelPRL2014}
\bibinfo{author}{Stannigel, K.} \emph{et~al.}
\newblock \bibinfo{title}{Constrained dynamics via the {Z}eno effect in quantum
  simulation: Implementing non-abelian lattice gauge theories with cold atoms}.
\newblock \emph{\bibinfo{journal}{Phys. Rev. Lett.}}
  \textbf{\bibinfo{volume}{112}}, \bibinfo{pages}{120406}
  (\bibinfo{year}{2014}).
\newblock \urlprefix\url{http://dx.doi.org/10.1103/PhysRevLett.112.120406}.

\bibitem{ZhuPRL2014}
\bibinfo{author}{Zhu, B.} \emph{et~al.}
\newblock \bibinfo{title}{Suppressing the loss of ultracold molecules via the
  continuous quantum {Z}eno effect}.
\newblock \emph{\bibinfo{journal}{Phys. Rev. Lett.}}
  \textbf{\bibinfo{volume}{112}}, \bibinfo{pages}{070404}
  (\bibinfo{year}{2014}).
\newblock \urlprefix\url{http://dx.doi.org/10.1103/PhysRevLett.112.070404}.

\bibitem{SchafferNatCommun2014}
\bibinfo{author}{Sch\"{a}ffer, F.} \emph{et~al.}
\newblock \bibinfo{title}{Experimental realization of quantum {Z}eno dynamics}.
\newblock \emph{\bibinfo{journal}{Nat. Commun.}} \textbf{\bibinfo{volume}{5}},
  \bibinfo{pages}{3194} (\bibinfo{year}{2014}).
\newblock \urlprefix\url{http://dx.doi.org/ 10.1038/ncomms4194}.

\bibitem{SignolesNaturePhysics2014}
\bibinfo{author}{{Signoles}, A.} \emph{et~al.}
\newblock \bibinfo{title}{{Confined quantum {Z}eno dynamics of a watched atomic
  arrow}}.
\newblock \emph{\bibinfo{journal}{Nat. Phys.}} \textbf{\bibinfo{volume}{10}},
  \bibinfo{pages}{715--719} (\bibinfo{year}{2014}).
\newblock \urlprefix\url{http://dx.doi.org/ 10.1038/nphys3076}.

\bibitem{DebierrePRA2015}
\bibinfo{author}{Debierre, V.}, \bibinfo{author}{Goessens, I.},
  \bibinfo{author}{Brainis, E.} \& \bibinfo{author}{Durt, T.}
\newblock \bibinfo{title}{Fermi's golden rule beyond the {Z}eno regime}.
\newblock \emph{\bibinfo{journal}{Phys. Rev. A}} \textbf{\bibinfo{volume}{92}},
  \bibinfo{pages}{023825} (\bibinfo{year}{2015}).
\newblock \urlprefix\url{http://dx.doi.org/doi/10.1103/PhysRevA.92.023825}.

\bibitem{AlexanderPRA2015}
\bibinfo{author}{Kiilerich, A.~H.} \& \bibinfo{author}{M\o{}lmer, K.}
\newblock \bibinfo{title}{Quantum {Z}eno effect in parameter estimation}.
\newblock \emph{\bibinfo{journal}{Phys. Rev. A}} \textbf{\bibinfo{volume}{92}},
  \bibinfo{pages}{032124} (\bibinfo{year}{2015}).
\newblock \urlprefix\url{http:///dx.doi.org/10.1103/PhysRevA.92.032124}.

\bibitem{QiuSciRep2015}
\bibinfo{author}{Qiu, J.} \emph{et~al.}
\newblock \bibinfo{title}{Quantum {Z}eno and {Z}eno-like effects in nitrogen
  vacancy centers}.
\newblock \emph{\bibinfo{journal}{Sci. Rep.}} \textbf{\bibinfo{volume}{5}},
  \bibinfo{pages}{17615} (\bibinfo{year}{2015}).
\newblock \urlprefix\url{http://dx.doi.org/10.1038/srep17615}.

\bibitem{HePRA2018}
\bibinfo{author}{He, S.}, \bibinfo{author}{Wang, C.}, \bibinfo{author}{Duan,
  L.-W.} \& \bibinfo{author}{Chen, Q.-H.}
\newblock \bibinfo{title}{Zeno effect of an open quantum system in the presence
  of $1/f$ noise}.
\newblock \emph{\bibinfo{journal}{Phys. Rev. A}} \textbf{\bibinfo{volume}{97}},
  \bibinfo{pages}{022108} (\bibinfo{year}{2018}).
\newblock \urlprefix\url{https://link.aps.org/doi/10.1103/PhysRevA.97.022108}.

\bibitem{HanggiNJP2018}
\bibinfo{author}{Magazzu, L.}, \bibinfo{author}{Talkner, P.} \&
  \bibinfo{author}{Hanggi, P.}
\newblock \bibinfo{title}{Quantum brownian motion under generalized position
  measurements: a converse {Z}eno scenario}.
\newblock \emph{\bibinfo{journal}{New J. Phys.}} \textbf{\bibinfo{volume}{20}},
  \bibinfo{pages}{033001} (\bibinfo{year}{2018}).

\bibitem{HePRA2019}
\bibinfo{author}{He, S.}, \bibinfo{author}{Duan, L.-W.}, \bibinfo{author}{Wang,
  C.} \& \bibinfo{author}{Chen, Q.-H.}
\newblock \bibinfo{title}{Quantum {Z}eno effect in a circuit-qed system}.
\newblock \emph{\bibinfo{journal}{Phys. Rev. A}} \textbf{\bibinfo{volume}{99}},
  \bibinfo{pages}{052101} (\bibinfo{year}{2019}).
\newblock \urlprefix\url{https://link.aps.org/doi/10.1103/PhysRevA.99.052101}.

\bibitem{MullerAnnPhys}
\bibinfo{author}{Müller, M.~M.}, \bibinfo{author}{Gherardini, S.} \&
  \bibinfo{author}{Caruso, F.}
\newblock \bibinfo{title}{Quantum {Z}eno dynamics through stochastic
  protocols}.
\newblock \emph{\bibinfo{journal}{Annalen der Physik}}
  \textbf{\bibinfo{volume}{529}}, \bibinfo{pages}{1600206}
  (\bibinfo{year}{2017}).

\bibitem{KurizkiNature2000}
\bibinfo{author}{Kofman, A.~G.} \& \bibinfo{author}{Kurizki, G.}
\newblock \bibinfo{title}{Acceleration of quantum decay processes by frequent
  observations}.
\newblock \emph{\bibinfo{journal}{Nature (London)}}
  \textbf{\bibinfo{volume}{405}}, \bibinfo{pages}{546} (\bibinfo{year}{2000}).
\newblock \urlprefix\url{http://dx.doi.org/10.1038/35014537}.

\bibitem{RaizenPRL2001}
\bibinfo{author}{Fischer, M.~C.}, \bibinfo{author}{Guti\'errez-Medina, B.} \&
  \bibinfo{author}{Raizen, M.~G.}
\newblock \bibinfo{title}{Observation of the quantum {Z}eno and anti-{Z}eno
  effects in an unstable system}.
\newblock \emph{\bibinfo{journal}{Phys. Rev. Lett.}}
  \textbf{\bibinfo{volume}{87}}, \bibinfo{pages}{040402}
  (\bibinfo{year}{2001}).
\newblock \urlprefix\url{http://dx.doi.org/10.1103/PhysRevLett.87.040402}.

\bibitem{BaronePRL2004}
\bibinfo{author}{Barone, A.}, \bibinfo{author}{Kurizki, G.} \&
  \bibinfo{author}{Kofman, A.~G.}
\newblock \bibinfo{title}{Dynamical control of macroscopic quantum tunneling}.
\newblock \emph{\bibinfo{journal}{Phys. Rev. Lett.}}
  \textbf{\bibinfo{volume}{92}}, \bibinfo{pages}{200403}
  (\bibinfo{year}{2004}).
\newblock \urlprefix\url{http://dx.doi.org/10.1103/PhysRevLett.92.200403}.

\bibitem{KoshinoPhysRep2005}
\bibinfo{author}{Koshino, K.} \& \bibinfo{author}{Shimizu, A.}
\newblock \bibinfo{title}{Quantum {Z}eno effect by general measurements}.
\newblock \emph{\bibinfo{journal}{Phys. Rep.}} \textbf{\bibinfo{volume}{412}},
  \bibinfo{pages}{191} (\bibinfo{year}{2005}).
\newblock \urlprefix\url{http://dx.doi.org/10.1088/1367-2630/18/5/053031}.

\bibitem{BennettPRB2010}
\bibinfo{author}{Chen, P.-W.}, \bibinfo{author}{Tsai, D.-B.} \&
  \bibinfo{author}{Bennett, P.}
\newblock \bibinfo{title}{Quantum {Z}eno and anti-{Z}eno effect of a
  nanomechanical resonator measured by a point contact}.
\newblock \emph{\bibinfo{journal}{Phys. Rev. B}} \textbf{\bibinfo{volume}{81}},
  \bibinfo{pages}{115307} (\bibinfo{year}{2010}).
\newblock \urlprefix\url{http://dx.doi.org/10.1103/PhysRevB.81.115307}.

\bibitem{YamamotoPRA2010}
\bibinfo{author}{Fujii, K.} \& \bibinfo{author}{Yamamoto, K.}
\newblock \bibinfo{title}{Anti-{Z}eno effect for quantum transport in
  disordered systems}.
\newblock \emph{\bibinfo{journal}{Phys. Rev. A}} \textbf{\bibinfo{volume}{82}},
  \bibinfo{pages}{042109} (\bibinfo{year}{2010}).
\newblock \urlprefix\url{http://dx.doi.org/10.1103/PhysRevA.82.042109}.

\bibitem{ChaudhryPRA2014zeno}
\bibinfo{author}{Chaudhry, A.~Z.} \& \bibinfo{author}{Gong, J.}
\newblock \bibinfo{title}{{Z}eno and anti-{Z}eno effects on dephasing}.
\newblock \emph{\bibinfo{journal}{Phys. Rev. A}} \textbf{\bibinfo{volume}{90}},
  \bibinfo{pages}{012101} (\bibinfo{year}{2014}).
\newblock \urlprefix\url{http://dx.doi.org/10.1103/PhysRevA.90.012101}.

\bibitem{Chaudhryscirep2017b}
\bibinfo{author}{Aftab, M.~J.} \& \bibinfo{author}{Chaudhry, A.~Z.}
\newblock \bibinfo{title}{Analyzing the quantum {Z}eno and anti-{Z}eno effects
  using optimal projective measurements}.
\newblock \emph{\bibinfo{journal}{Sci. Rep.}} \textbf{\bibinfo{volume}{7}},
  \bibinfo{pages}{11766} (\bibinfo{year}{2017}).

\bibitem{HePRA2017}
\bibinfo{author}{He, S.}, \bibinfo{author}{Chen, Q.-H.} \&
  \bibinfo{author}{Zheng, H.}
\newblock \bibinfo{title}{Zeno and anti-{Z}eno effect in an open quantum system
  in the ultrastrong-coupling regime}.
\newblock \emph{\bibinfo{journal}{Phys. Rev. A}} \textbf{\bibinfo{volume}{95}},
  \bibinfo{pages}{062109} (\bibinfo{year}{2017}).
\newblock \urlprefix\url{https://link.aps.org/doi/10.1103/PhysRevA.95.062109}.

\bibitem{WuPRA2017}
\bibinfo{author}{Wu, W.} \& \bibinfo{author}{Lin, H.-Q.}
\newblock \bibinfo{title}{Quantum {Z}eno and anti-{Z}eno effects in quantum
  dissipative systems}.
\newblock \emph{\bibinfo{journal}{Phys. Rev. A}} \textbf{\bibinfo{volume}{95}},
  \bibinfo{pages}{042132} (\bibinfo{year}{2017}).
\newblock \urlprefix\url{http://dx.doi.org/10.1103/PhysRevA.95.042132}.

\bibitem{Chaudhryscirep2018}
\bibinfo{author}{Majeed, M.} \& \bibinfo{author}{Chaudhry, A.~Z.}
\newblock \bibinfo{title}{The quantum {Z}eno and anti-{Z}eno effects with
  non-selective projective measurements}.
\newblock \emph{\bibinfo{journal}{Sci. Rep.}} \textbf{\bibinfo{volume}{8}},
  \bibinfo{pages}{14887} (\bibinfo{year}{2018}).

\bibitem{WuAnnals2018}
\bibinfo{author}{Wu, W.}
\newblock \bibinfo{title}{Quantum {Z}eno and anti-{Z}eno dynamics in a spin
  environment}.
\newblock \emph{\bibinfo{journal}{Ann.~Phys.}} \textbf{\bibinfo{volume}{396}},
  \bibinfo{pages}{147} (\bibinfo{year}{2018}).

\bibitem{ChaudhryEJPD2019a}
\bibinfo{author}{Khalid, B.} \& \bibinfo{author}{Chaudhry, A.~Z.}
\newblock \bibinfo{title}{The quantum {Z}eno and anti-{Z}eno effects: from weak
  to strong system-environment coupling}.
\newblock \emph{\bibinfo{journal}{Eur. J. Phys. D}}
  \textbf{\bibinfo{volume}{73}}, \bibinfo{pages}{134} (\bibinfo{year}{2019}).

\bibitem{sakuldee2020}
\bibinfo{author}{Sakuldee, F.} \& \bibinfo{author}{Cywi{\'n}ski, {\L}.}
\newblock \bibinfo{title}{Spectroscopy of classical environmental noise with a
  qubit subjected to projective measurements}.
\newblock \emph{\bibinfo{journal}{Phys. Rev. A}}
  \textbf{\bibinfo{volume}{101}}, \bibinfo{pages}{012314}
  (\bibinfo{year}{2020}).
\newblock \urlprefix\url{http://dx.doi.org/ 10.1103/PhysRevA.101.012314}.

\bibitem{mullerPLA2020}
\bibinfo{author}{M{\"u}ller, M.~M.}, \bibinfo{author}{Gherardini, S.},
  \bibinfo{author}{Dalla~Pozza, N.} \& \bibinfo{author}{Caruso, F.}
\newblock \bibinfo{title}{Noise sensing via stochastic quantum {Z}eno}.
\newblock \emph{\bibinfo{journal}{Phys. Lett. A}}
  \textbf{\bibinfo{volume}{384}}, \bibinfo{pages}{126244}
  (\bibinfo{year}{2020}).
\newblock \urlprefix\url{http://dx.doi.org/10.1016/j.physleta.2020.126244}.

\bibitem{sakuldeePRA2020}
\bibinfo{author}{Sakuldee, F.} \& \bibinfo{author}{Cywi{\'n}ski, {\L}.}
\newblock \bibinfo{title}{Relationship between subjecting the qubit to
  dynamical decoupling and to a sequence of projective measurements}.
\newblock \emph{\bibinfo{journal}{Phys. Rev. A}}
  \textbf{\bibinfo{volume}{101}}, \bibinfo{pages}{042329}
  (\bibinfo{year}{2020}).
\newblock \urlprefix\url{http://dx.doi.org/ 10.1103/PhysRevA.101.042329}.

\bibitem{ManiscalcoPRL2006}
\bibinfo{author}{Maniscalco, S.}, \bibinfo{author}{Piilo, J.} \&
  \bibinfo{author}{Suominen, K.-A.}
\newblock \bibinfo{title}{{Z}eno and anti-{Z}eno effects for quantum brownian
  motion}.
\newblock \emph{\bibinfo{journal}{Phys. Rev. Lett.}}
  \textbf{\bibinfo{volume}{97}}, \bibinfo{pages}{130402}
  (\bibinfo{year}{2006}).
\newblock \urlprefix\url{http://dx.doi.org/10.1103/PhysRevLett.97.130402}.

\bibitem{SegalPRA2007}
\bibinfo{author}{Segal, D.} \& \bibinfo{author}{Reichman, D.~R.}
\newblock \bibinfo{title}{{Z}eno and anti-{Z}eno effects in spin-bath models}.
\newblock \emph{\bibinfo{journal}{Phys. Rev. A}} \textbf{\bibinfo{volume}{76}},
  \bibinfo{pages}{012109} (\bibinfo{year}{2007}).
\newblock \urlprefix\url{http://dx.doi.org/10.1103/PhysRevA.76.012109}.

\bibitem{ZhengPRL2008}
\bibinfo{author}{Zheng, H.}, \bibinfo{author}{Zhu, S.~Y.} \&
  \bibinfo{author}{Zubairy, M.~S.}
\newblock \bibinfo{title}{Quantum {Z}eno and anti-{Z}eno effects: Without the
  rotating-wave approximation}.
\newblock \emph{\bibinfo{journal}{Phys. Rev. Lett.}}
  \textbf{\bibinfo{volume}{101}}, \bibinfo{pages}{200404}
  (\bibinfo{year}{2008}).
\newblock \urlprefix\url{http://dx.doi.org/10.1103/PhysRevLett.101.200404}.

\bibitem{AiPRA2010}
\bibinfo{author}{Ai, Q.}, \bibinfo{author}{Li, Y.}, \bibinfo{author}{Zheng, H.}
  \& \bibinfo{author}{Sun, C.~P.}
\newblock \bibinfo{title}{Quantum anti-{Z}eno effect without rotating wave
  approximation}.
\newblock \emph{\bibinfo{journal}{Phys. Rev. A}} \textbf{\bibinfo{volume}{81}},
  \bibinfo{pages}{042116} (\bibinfo{year}{2010}).
\newblock \urlprefix\url{http://dx.doi.org/10.1103/PhysRevA.81.042116}.

\bibitem{ThilagamJMP2010}
\bibinfo{author}{Thilagam, A.}
\newblock \bibinfo{title}{{Z}eno--anti-{Z}eno crossover dynamics in a
  spin--boson system}.
\newblock \emph{\bibinfo{journal}{J. Phys. A: Math. Theor.}}
  \textbf{\bibinfo{volume}{43}}, \bibinfo{pages}{155301}
  (\bibinfo{year}{2010}).
\newblock \urlprefix\url{http://dx.doi.org/ 10.1088/1751- 8113/43/15/155301}.

\bibitem{ThilagamJCP2013}
\bibinfo{author}{Thilagam, A.}
\newblock \bibinfo{title}{Non-markovianity during the quantum {Z}eno effect}.
\newblock \emph{\bibinfo{journal}{J. Chem. Phys.}}
  \textbf{\bibinfo{volume}{138}}, \bibinfo{pages}{175102}
  (\bibinfo{year}{2013}).
\newblock \urlprefix\url{http://dx.doi.org/ 10.1063/1.4802785}.

\bibitem{Chaudhryscirep2016}
\bibinfo{author}{Chaudhry, A.~Z.}
\newblock \bibinfo{title}{A general framework for the quantum {Z}eno and
  anti-{Z}eno effects}.
\newblock \emph{\bibinfo{journal}{Sci. Rep.}} \textbf{\bibinfo{volume}{6}},
  \bibinfo{pages}{29497} (\bibinfo{year}{2016}).
\newblock \urlprefix\url{http://dx.doi.org/10.1038/srep29497}.

\bibitem{Chaudhryscirep2017a}
\bibinfo{author}{Chaudhry, A.~Z.}
\newblock \bibinfo{title}{The quantum {Z}eno and anti-{Z}eno effects with
  strong system-environment coupling}.
\newblock \emph{\bibinfo{journal}{Sci. Rep.}} \textbf{\bibinfo{volume}{7}},
  \bibinfo{pages}{1741} (\bibinfo{year}{2017}).
\newblock \urlprefix\url{http://dx.doi.org/10.1038/s41598-017-01844-8}.

\bibitem{Hanggidrivenquantumtunneling}
\bibinfo{author}{Grifoni, M.} \& \bibinfo{author}{H{\"a}nggi, P.}
\newblock \bibinfo{title}{Driven quantum tunneling}.
\newblock \emph{\bibinfo{journal}{Phys. Rep.}} \textbf{\bibinfo{volume}{304}},
  \bibinfo{pages}{229--354} (\bibinfo{year}{1998}).
\newblock \urlprefix\url{http://dx.doi.org/10.1016/S0370-1573(98)00022-2}.

\bibitem{kofmanPRL2001}
\bibinfo{author}{Kofman, A.} \& \bibinfo{author}{Kurizki, G.}
\newblock \bibinfo{title}{Universal dynamical control of quantum mechanical
  decay: modulation of the coupling to the continuum}.
\newblock \emph{\bibinfo{journal}{Phys. Rev Lett.}}
  \textbf{\bibinfo{volume}{87}}, \bibinfo{pages}{270405}
  (\bibinfo{year}{2001}).
\newblock \urlprefix\url{http://dx.doi.org/10.1103/PhysRevLett.87.270405}.

\bibitem{kofmanPRL2004}
\bibinfo{author}{Kofman, A.} \& \bibinfo{author}{Kurizki, G.}
\newblock \bibinfo{title}{Unified theory of dynamically suppressed qubit
  decoherence in thermal baths}.
\newblock \emph{\bibinfo{journal}{Phys. Rev. Lett.}}
  \textbf{\bibinfo{volume}{93}}, \bibinfo{pages}{130406}
  (\bibinfo{year}{2004}).
\newblock \urlprefix\url{http://dx.doi.org/10.1103/PhysRevLett.93.130406}.

\bibitem{gordonJPB2007}
\bibinfo{author}{Gordon, G.}, \bibinfo{author}{Erez, N.} \&
  \bibinfo{author}{Kurizki, G.}
\newblock \bibinfo{title}{Universal dynamical decoherence control of noisy
  single-and multi-qubit systems}.
\newblock \emph{\bibinfo{journal}{J. Phys. B}} \textbf{\bibinfo{volume}{40}},
  \bibinfo{pages}{S75} (\bibinfo{year}{2007}).
\newblock \urlprefix\url{http://dx.doi.org/10.1088/0953-4075/40/9/S04}.

\bibitem{gordonPRL2008}
\bibinfo{author}{Gordon, G.}, \bibinfo{author}{Kurizki, G.} \&
  \bibinfo{author}{Lidar, D.~A.}
\newblock \bibinfo{title}{Optimal dynamical decoherence control of a qubit}.
\newblock \emph{\bibinfo{journal}{Phys. Rev. Lett.}}
  \textbf{\bibinfo{volume}{101}}, \bibinfo{pages}{010403}
  (\bibinfo{year}{2008}).
\newblock \urlprefix\url{http://dx.doi.org/ 10.1103/PhysRevLett.101.010403}.

\bibitem{noelPRA1998}
\bibinfo{author}{Noel, M.~W.}, \bibinfo{author}{Griffith, W.} \&
  \bibinfo{author}{Gallagher, T.}
\newblock \bibinfo{title}{Frequency-modulated excitation of a two-level atom}.
\newblock \emph{\bibinfo{journal}{Phys. Rev. A}} \textbf{\bibinfo{volume}{58}},
  \bibinfo{pages}{2265} (\bibinfo{year}{1998}).
\newblock \urlprefix\url{http://dx.doi.org/10.1103/PhysRevA.58.2265}.

\bibitem{grossmannPRL1991}
\bibinfo{author}{Grossmann, F.}, \bibinfo{author}{Dittrich, T.},
  \bibinfo{author}{Jung, P.} \& \bibinfo{author}{H{\"a}nggi, P.}
\newblock \bibinfo{title}{Coherent destruction of tunneling}.
\newblock \emph{\bibinfo{journal}{Phys. Rev. lett.}}
  \textbf{\bibinfo{volume}{67}}, \bibinfo{pages}{516} (\bibinfo{year}{1991}).
\newblock \urlprefix\url{http://dx.doi.org/10.1103/PhysRevLett.67.516}.

\bibitem{shaoPRA1997}
\bibinfo{author}{Shao, J.} \& \bibinfo{author}{H{\"a}nggi, P.}
\newblock \bibinfo{title}{Controlling quantum coherence by circularly polarized
  fields}.
\newblock \emph{\bibinfo{journal}{Phys. Rev. A}} \textbf{\bibinfo{volume}{56}},
  \bibinfo{pages}{R4397} (\bibinfo{year}{1997}).
\newblock \urlprefix\url{http://dx.doi.org/10.1103/PhysRevA.56.R4397}.

\bibitem{ViolaPRA1998}
\bibinfo{author}{Viola, L.} \& \bibinfo{author}{Lloyd, S.}
\newblock \bibinfo{title}{Dynamical suppression of decoherence in two-state
  quantum systems}.
\newblock \emph{\bibinfo{journal}{Phys. Rev. A}} \textbf{\bibinfo{volume}{58}},
  \bibinfo{pages}{2733--2744} (\bibinfo{year}{1998}).
\newblock \urlprefix\url{http://link.aps.org/doi/10.1103/PhysRevA.58.2733}.

\bibitem{LloydPRL1999}
\bibinfo{author}{Viola, L.}, \bibinfo{author}{Knill, E.} \&
  \bibinfo{author}{Lloyd, S.}
\newblock \bibinfo{title}{Dynamical decoupling of open quantum systems}.
\newblock \emph{\bibinfo{journal}{Phys. Rev. Lett.}}
  \textbf{\bibinfo{volume}{82}}, \bibinfo{pages}{2417--2421}
  (\bibinfo{year}{1999}).
\newblock \urlprefix\url{http://link.aps.org/doi/10.1103/PhysRevLett.82.2417}.

\bibitem{FanchiniPRA12007}
\bibinfo{author}{Fanchini, F.~F.}, \bibinfo{author}{Hornos, J. E.~M.} \&
  \bibinfo{author}{Napolitano, R. d.~J.}
\newblock \bibinfo{title}{Continuously decoupling single-qubit operations from
  a perturbing thermal bath of scalar bosons}.
\newblock \emph{\bibinfo{journal}{Phys. Rev. A}} \textbf{\bibinfo{volume}{75}},
  \bibinfo{pages}{022329} (\bibinfo{year}{2007}).
\newblock \urlprefix\url{http://link.aps.org/doi/10.1103/PhysRevA.75.022329}.

\bibitem{ChaudhryPRA12012}
\bibinfo{author}{Chaudhry, A.~Z.} \& \bibinfo{author}{Gong, J.}
\newblock \bibinfo{title}{Decoherence control: Universal protection of
  two-qubit states and two-qubit gates using continuous driving fields}.
\newblock \emph{\bibinfo{journal}{Phys. Rev. A}} \textbf{\bibinfo{volume}{85}},
  \bibinfo{pages}{012315} (\bibinfo{year}{2012}).
\newblock \urlprefix\url{http://link.aps.org/doi/10.1103/PhysRevA.85.012315}.

\bibitem{ChaudhryPRA22012}
\bibinfo{author}{Chaudhry, A.~Z.} \& \bibinfo{author}{Gong, J.}
\newblock \bibinfo{title}{Protecting and enhancing spin squeezing via
  continuous dynamical decoupling}.
\newblock \emph{\bibinfo{journal}{Phys. Rev. A}} \textbf{\bibinfo{volume}{86}},
  \bibinfo{pages}{012311} (\bibinfo{year}{2012}).
\newblock \urlprefix\url{http://link.aps.org/doi/10.1103/PhysRevA.86.012311}.

\bibitem{ChaudhryPRA2019}
\bibinfo{author}{Austin, S.}, \bibinfo{author}{Khan, M.~Q.},
  \bibinfo{author}{Mudassar, M.} \& \bibinfo{author}{Chaudhry, A.~Z.}
\newblock \bibinfo{title}{Continuous dynamical decoupling of spin chains:
  Modulating the spin-environment and spin-spin interactions}.
\newblock \emph{\bibinfo{journal}{Phys. Rev. A}}
  \textbf{\bibinfo{volume}{100}}, \bibinfo{pages}{022102}
  (\bibinfo{year}{2019}).
\newblock \urlprefix\url{https://link.aps.org/doi/10.1103/PhysRevA.100.022102}.

\bibitem{doNJP2019}
\bibinfo{author}{Do, H.-V.} \emph{et~al.}
\newblock \bibinfo{title}{Experimental proof of quantum {Z}eno-assisted noise
  sensing}.
\newblock \emph{\bibinfo{journal}{New J. Phys.}} \textbf{\bibinfo{volume}{21}},
  \bibinfo{pages}{113056} (\bibinfo{year}{2019}).
\newblock \urlprefix\url{http://dx.doi.org/10.1088/1367-2630/ab5740}.

\bibitem{VorrathPRL2005}
\bibinfo{author}{Vorrath, T.} \& \bibinfo{author}{Brandes, T.}
\newblock \bibinfo{title}{Dynamics of a large spin with strong dissipation}.
\newblock \emph{\bibinfo{journal}{Phys. Rev. Lett.}}
  \textbf{\bibinfo{volume}{95}}, \bibinfo{pages}{070402}
  (\bibinfo{year}{2005}).
\newblock \urlprefix\url{http://dx.doi.org/10.1103/PhysRevLett.95.070402}.

\bibitem{SilbeyJCP1984}
\bibinfo{author}{Silbey, R.} \& \bibinfo{author}{Harris, R.~A.}
\newblock \bibinfo{title}{Variational calculation of the dynamics of a two
  level system interacting with a bath}.
\newblock \emph{\bibinfo{journal}{J. Chem. Phys.}}
  \textbf{\bibinfo{volume}{80}}, \bibinfo{pages}{2615--2617}
  (\bibinfo{year}{1984}).
\newblock \urlprefix\url{http://dx.doi.org/10.1063/1.447055}.

\bibitem{Vorraththesis}
\bibinfo{author}{Vorrath, T.}
\newblock \emph{\bibinfo{title}{Dissipation-Induced Collective Effects in
  Two-Level Systems}}.
\newblock Ph.D. thesis (\bibinfo{year}{2003}).

\bibitem{LeeJCP2012}
\bibinfo{author}{Lee, C.~K.}, \bibinfo{author}{Moix, J.} \&
  \bibinfo{author}{Cao, J.}
\newblock \bibinfo{title}{Accuracy of second order perturbation theory in the
  polaron and variational polaron frames}.
\newblock \emph{\bibinfo{journal}{J. Chem. Phys.}}
  \textbf{\bibinfo{volume}{136}}, \bibinfo{pages}{204120}
  (\bibinfo{year}{2012}).
\newblock \urlprefix\url{http://dx.doi.org/10.1103/PhysRevE.86.021109}.

\bibitem{changJCP2013}
\bibinfo{author}{Chang, H.-T.}, \bibinfo{author}{Zhang, P.-P.} \&
  \bibinfo{author}{Cheng, Y.-C.}
\newblock \bibinfo{title}{Criteria for the accuracy of small polaron quantum
  master equation in simulating excitation energy transfer dynamics}.
\newblock \emph{\bibinfo{journal}{J. of Chem. Phys.}}
  \textbf{\bibinfo{volume}{139}}, \bibinfo{pages}{224112}
  (\bibinfo{year}{2013}).
\newblock \urlprefix\url{http://dx.doi.org/10.1063/1.4840795}.

\bibitem{jang2008theory}
\bibinfo{author}{Jang, S.}, \bibinfo{author}{Cheng, Y.-C.},
  \bibinfo{author}{Reichman, D.~R.} \& \bibinfo{author}{Eaves, J.~D.}
\newblock \bibinfo{title}{Theory of coherent resonance energy transfer}.
\newblock \emph{\bibinfo{journal}{J. Chem. Phys.}}
  \textbf{\bibinfo{volume}{129}}, \bibinfo{pages}{101104}
  (\bibinfo{year}{2008}).
\newblock \urlprefix\url{http://dx.doi.org/10.1063/1.2977974}.

\bibitem{ChinPRL2011}
\bibinfo{author}{Chin, A.~W.}, \bibinfo{author}{Prior, J.},
  \bibinfo{author}{Huelga, S.~F.} \& \bibinfo{author}{Plenio, M.~B.}
\newblock \bibinfo{title}{Generalized polaron ansatz for the ground state of
  the sub-ohmic spin-boson model: An analytic theory of the localization
  transition}.
\newblock \emph{\bibinfo{journal}{Phys. Rev. Lett.}}
  \textbf{\bibinfo{volume}{107}}, \bibinfo{pages}{160601}
  (\bibinfo{year}{2011}).
\newblock \urlprefix\url{http://dx.doi.org/10.1063/1.4722336}.

\bibitem{GuzikJPCL2015}
\bibinfo{author}{Gelbwaser-Klimovsky, D.} \& \bibinfo{author}{Aspuru-Guzik, A.}
\newblock \bibinfo{title}{Strongly coupled quantum heat machines}.
\newblock \emph{\bibinfo{journal}{J. Chem. Phys. Lett.}}
  \textbf{\bibinfo{volume}{6}}, \bibinfo{pages}{3477--3482}
  (\bibinfo{year}{2015}).
\newblock \urlprefix\url{http://dx.doi.org/10.1021/acs.jpclett.5b01404}.

\bibitem{Sakuraibook}
\bibinfo{author}{Sakurai, J.~J.}
\newblock \emph{\bibinfo{title}{Modern Quantum Mechanics}}
  (\bibinfo{publisher}{Addison Wesley}, \bibinfo{address}{Reading, Mass.},
  \bibinfo{year}{1993}).

\bibitem{MilitelloPRA2019a}
\bibinfo{author}{Militello, B.}
\newblock \bibinfo{title}{Three-state landau-zener model in the presence of
  dissipation}.
\newblock \emph{\bibinfo{journal}{Phys. Rev. A}} \textbf{\bibinfo{volume}{99}},
  \bibinfo{pages}{033415} (\bibinfo{year}{2019}).
\newblock \urlprefix\url{https://link.aps.org/doi/10.1103/PhysRevA.99.033415}.

\bibitem{MilitelloPRA2019b}
\bibinfo{author}{Militello, B.}
\newblock \bibinfo{title}{Detuning-induced robustness of a three-state
  landau-zener model against dissipation}.
\newblock \emph{\bibinfo{journal}{Phys. Rev. A}} \textbf{\bibinfo{volume}{99}},
  \bibinfo{pages}{063412} (\bibinfo{year}{2019}).
\newblock \urlprefix\url{https://link.aps.org/doi/10.1103/PhysRevA.99.063412}.

\bibitem{Halliwellhistoriesreview}
\bibinfo{author}{Halliwell, J.}
\newblock \bibinfo{title}{A review of the decoherent histories approach to
  quantum mechanics}.
\newblock \emph{\bibinfo{journal}{Annals N.Y. Acad. Sci.}}
  \textbf{\bibinfo{volume}{755}}, \bibinfo{pages}{726} (\bibinfo{year}{1995}).

\bibitem{DankoPRA2018}
\bibinfo{author}{Georgiev, D.} \& \bibinfo{author}{Cohen, E.}
\newblock \bibinfo{title}{Probing finite coarse-grained virtual feynman
  histories with sequential weak values}.
\newblock \emph{\bibinfo{journal}{Phys. Rev. A}} \textbf{\bibinfo{volume}{97}},
  \bibinfo{pages}{052102} (\bibinfo{year}{2018}).
\newblock \urlprefix\url{https://link.aps.org/doi/10.1103/PhysRevA.97.052102}.

\bibitem{FacchiPRL2001}
\bibinfo{author}{Facchi, P.}, \bibinfo{author}{Nakazato, H.} \&
  \bibinfo{author}{Pascazio, S.}
\newblock \bibinfo{title}{From the quantum {Z}eno to the inverse quantum {Z}eno
  effect}.
\newblock \emph{\bibinfo{journal}{Phys. Rev. Lett.}}
  \textbf{\bibinfo{volume}{86}}, \bibinfo{pages}{2699--2703}
  (\bibinfo{year}{2001}).
\newblock \urlprefix\url{https://link.aps.org/doi/10.1103/PhysRevLett.86.2699}.

\bibitem{VegaRMP2017}
\bibinfo{author}{de~Vega, I.} \& \bibinfo{author}{Alonso, D.}
\newblock \bibinfo{title}{Dynamics of non-markovian open quantum systems}.
\newblock \emph{\bibinfo{journal}{Rev. Mod. Phys.}}
  \textbf{\bibinfo{volume}{89}}, \bibinfo{pages}{015001}
  (\bibinfo{year}{2017}).

\bibitem{Gradshteynbook}
\bibinfo{author}{Gradshteyn, I.~M.} \& \bibinfo{author}{Ryzhik, I.~S.}
\newblock \emph{\bibinfo{title}{Table of Integrals, Series, and Products}}
  (\bibinfo{publisher}{Academic Press}, \bibinfo{address}{San Diego},
  \bibinfo{year}{1994}).

\bibitem{KurizkiPRL2011}
\bibinfo{author}{Bar-Gill, N.}, \bibinfo{author}{Rao, D. D.~B.} \&
  \bibinfo{author}{Kurizki, G.}
\newblock \bibinfo{title}{Creating nonclassical states of bose-einstein
  condensates by dephasing collisions}.
\newblock \emph{\bibinfo{journal}{Phys. Rev. Lett.}}
  \textbf{\bibinfo{volume}{107}}, \bibinfo{pages}{010404}
  (\bibinfo{year}{2011}).

\bibitem{Dicke1954}
\bibinfo{author}{Dicke, R.~H.}
\newblock \bibinfo{title}{Coherence in spontaneous radiation processes}.
\newblock \emph{\bibinfo{journal}{Phys. Rev.}} \textbf{\bibinfo{volume}{93}},
  \bibinfo{pages}{99--110} (\bibinfo{year}{1954}).
\newblock \urlprefix\url{http://link.aps.org/doi/10.1103/PhysRev.93.99}.

\bibitem{PollakPRE2008}
\bibinfo{author}{Pollak, E.}, \bibinfo{author}{Shao, J.} \&
  \bibinfo{author}{Zhang, D.~H.}
\newblock \bibinfo{title}{Effects of initial correlations on the dynamics of
  dissipative systems}.
\newblock \emph{\bibinfo{journal}{Phys. Rev. E}} \textbf{\bibinfo{volume}{77}},
  \bibinfo{pages}{021107} (\bibinfo{year}{2008}).
\newblock \urlprefix\url{http://link.aps.org/doi/10.1103/PhysRevE.77.021107}.

\bibitem{ChaudhryPRA2013a}
\bibinfo{author}{Chaudhry, A.~Z.} \& \bibinfo{author}{Gong, J.}
\newblock \bibinfo{title}{Amplification and suppression of
  system-bath-correlation effects in an open many-body system}.
\newblock \emph{\bibinfo{journal}{Phys. Rev. A}} \textbf{\bibinfo{volume}{87}},
  \bibinfo{pages}{012129} (\bibinfo{year}{2013}).

\bibitem{ChaudhryPRA2013b}
\bibinfo{author}{Chaudhry, A.~Z.} \& \bibinfo{author}{Gong, J.}
\newblock \bibinfo{title}{Role of initial system-environment correlations: A
  master equation approach}.
\newblock \emph{\bibinfo{journal}{Phys. Rev. A}} \textbf{\bibinfo{volume}{88}},
  \bibinfo{pages}{052107} (\bibinfo{year}{2013}).

\bibitem{ChaudhryEJPD2019b}
\bibinfo{author}{Majeed, M.} \& \bibinfo{author}{Chaudhry, A.~Z.}
\newblock \bibinfo{title}{Effect of initial system–environment correlations
  with spin environments}.
\newblock \emph{\bibinfo{journal}{Eur. J. Phys. D}}
  \textbf{\bibinfo{volume}{73}}, \bibinfo{pages}{1} (\bibinfo{year}{2019}).

\bibitem{ChaudhryPRA2020}
\bibinfo{author}{Austin, S.}, \bibinfo{author}{Zahid, S.} \&
  \bibinfo{author}{Chaudhry, A.~Z.}
\newblock \bibinfo{title}{Geometric phase corrected by initial
  system-environment correlations}.
\newblock \emph{\bibinfo{journal}{Phys. Rev. A}}
  \textbf{\bibinfo{volume}{101}}, \bibinfo{pages}{022114}
  (\bibinfo{year}{2020}).
\newblock \urlprefix\url{https://link.aps.org/doi/10.1103/PhysRevA.101.022114}.

\bibitem{Kurizki2015}
\bibinfo{author}{Kurizki, G.}, \bibinfo{author}{Shahmoon, E.} \&
  \bibinfo{author}{Zwick, A.}
\newblock \bibinfo{title}{Thermal baths as quantum resources: more friends than
  foes?}
\newblock \emph{\bibinfo{journal}{Physica Scripta}}
  \textbf{\bibinfo{volume}{90}}, \bibinfo{pages}{128002}
  (\bibinfo{year}{2015}).
\newblock
  \urlprefix\url{https://doi.org/10.1088%2F0031-8949%2F90%2F12%2F128002}.

\end{thebibliography}
\end{document}